\documentclass[fleqn,usenatbib]{mnras}

\usepackage{newtxtext,newtxmath}
\usepackage[T1]{fontenc}
\usepackage{ae,aecompl}

\usepackage{graphicx}	% Including figure files
\usepackage{amsmath}	% Advanced maths commands
\usepackage{amssymb}	% Extra maths symbols
\usepackage{color}

\title[Early LC of Type II SNe interacting with a CS disk]{Early light curves of Type II supernovae interacting with a circumstellar disk}
\author[T. Nagao et al.]{
T. Nagao,$^{1,2}$\thanks{E-mail: Takashi.Nagao@eso.org}
K. Maeda,$^{2}$
and R. Ouchi,$^{2}$
\\
$^{1}$European Southern Observatory, Karl-Schwarzschild-Str. 2, 85748 Garching b. M\"{u}nchen, Germany\\
$^{2}$Department of astronomy, Kyoto University, Kitashirakawa-Oiwake-cho,
  Sakyo-ku, Kyoto 606-8502, Japan\\
}

\date{Accepted XXX. Received YYY; in original form ZZZ}

\pubyear{2020}

\begin{document}
\label{firstpage}
\pagerange{\pageref{firstpage}--\pageref{lastpage}}
\maketitle

\begin{abstract}
Type II supernovae (SNe) interacting with disklike circumstellar matter (CSM) have been suggested as an explanation of some unusual Type II SNe, e.g., the so-called ``impossible'' SN, iPTF14hls. There are some radiation hydrodynamics simulations for such SNe interacting with a CSM disk. However, such disk interaction models so far have not included the effect of the ionization and recombination processes in the SN ejecta, i.e., the fact that the photosphere of Type IIP SNe between $\sim 10$-$\sim 100$ days is regulated by the hydrogen recombination front.
We calculate light curves for Type IIP SNe interacting with a CSM disk viewed from the polar direction, and examine the effects of the disk density and opening angle on their bolometric light curves. This work embeds the shock interaction model of \citet{Moriya2013} within the Type IIP SN model of \citet{Kasen2009}, for taking into account the effects of the ionization and recombination in the SN ejecta.
We demonstrate that such interacting SNe show three phases with different photometric and spectroscopic properties, following the change in the energy source: First few tens days after explosion (Phase~1), $\sim 10 - \sim 100$ days (Phase~2) and days after that (Phase~3).
%(Phase~1): For first few tens days, the main energy source is the thermal energy of the SN ejecta deposited by the core-bounced shock and the synthesized radioactive elements. The energy released by the CSM interaction is injected deep into the SN ejecta, most of which is lost by the expansion cooling within the SN ejecta. Thus, the system shows similar photospheric and spectroscopic properties with those of canonical Type II SNe with P-Cygni lines. 
%(Phase~2): Once the CSM interaction overtakes the photosphere in the SN ejecta, the interacting region becomes exposed to an observer. Then, the radiation from the system becomes a mixture of the radiation from the SN ejecta and the CSM interaction, showing high luminosity and some interaction features in the spectrum such as narrow emission lines. 
%(Phase~3): Finally, the system becomes, $\sim 100$ days after explosion, powered by the CSM interaction and the radioactive decay because of the exhaustion of the initial thermal energy in the SN ejecta. 
%
From the calculations, we conclude that such hidden CSM disk cannot account for overluminous Type IIP SNe. We find that the luminosity ratio between Phase~1 and Phase~2 has information on the opening angle of the CSM disk. We thus encourage early photometric and spectroscopic observations of interacting SNe for investigating their CSM geometry.
\end{abstract}

\begin{keywords}
supernovae: general -- circumstellar matter -- stars: mass-loss
\end{keywords}

%%%%%%%%%%%%%%%%%%%%%%%%%%%%%%%%%%%%%%%%%%%%%%%%%%

%%%%%%%%%%%%%%%%% BODY OF PAPER %%%%%%%%%%%%%%%%%%

%%%%%%%%%%%%%%%%% INTRODUCTION %%%%%%%%%%%%%%%%%%
\section{Introduction}

The understanding of the final evolutionary stage of massive stars is one of the biggest problems in the theory of stellar evolution. Evidence on the mass eruption at the final evolutionary stage of massive stars has been steadily accumulating over the years, while its origin has not been clarified yet. Historically such mysterious mass loss has been recognized by observations of Type IIn supernovae (SNe), which are believed to shine mainly by an interaction between the SN ejecta and circumstellar matter (CSM). Since CSM in the vicinity of SNe should be created by mass loss just before explosion, Type IIn SNe demonstrate the existence of huge mass loss just before explosion ($\sim 10^{-4} - \sim 1$ M$_{\odot}$ yr$^{-1}$; e.g., \citealt{Smith2017} for a recent review; see also \citealt{Filippenko1997, Kiewe2012, Taddia2013}). The estimated amount and time-scale for their mass-loss from the light-curve analysis are very diverse. In the case of SN~2006gy, the CSM is thought to originate from an eruption that ejected $\sim 20$ M$_{\odot}$ $\sim 8$ years before the explosion \citep[][but see \cite{Jerkstrand2020}]{Smith2010}, while the CSM of SN~2010jl is thought to be created by a relatively steady mass loss that lasts at least for several hundreds years before the explosion with a mass-loss rate of $\gtrsim 10^{-1}$ M$_{\odot}$ yr$^{-1}$ \citep[][]{Fransson2014}.
There is a growing number of Type IIn SNe whose progenitors show photometric variability just before exploding as an SN \citep[e.g.,][]{Fraser2013, Mauerhan2013, Pastorello2013, Ofek2014, Elias-Rosa2016, Thone2017, Kilpatrick2018, Pastorello2018, Pastorello2019}. This photometric variability is regarded as the direct evidence of some eruptive mass loss from massive stars in the very last stage of their lives.
Recently, it has also been claimed that massive stars generally experience huge mass loss several years before explosion ($\sim 10^{-4} - \sim 1$ M$_{\odot}$ yr$^{-1}$), from early-phase observations of SN spectra \citep[e.g.,][]{Khazov2016, Yaron2017, Boian2020} and light curves \citep[e.g.,][]{Forster2018}.

The mechanism of such huge mass loss is still unclear, although some mechanisms have been proposed such as outbursts caused by some stellar instabilities \citep[e.g.,][]{Humphreys1994, Langer1999, Yoon2010, Arnett2011, Quataert2012, Shiode2014, Smith2014, Woosley2015, Quataert2016, Fuller2017}, or mass loss due to binary interaction \citep[e.g.,][]{Chevalier2012, Soker2013}.

It is important to investigate the spacial distribution of CSM for interacting SNe, which is directly related with the unknown mass-loss mechanism. In fact, some aspherical CSM geometries have been implied by polarimetric \citep[e.g.,][]{Leonard2000, Wang2001, Hoffman2008, Patat2011, Mauerhan2014, Reilly2017}, spectroscopic \citep[e.g.,][]{Fransson2002, Levesque2014, Fransson2014, Andrews2017}, ultraviolet \citep[e.g.,][]{Soumagnac2019, Soumagnac2020}, and X-ray \citep[e.g.,][]{Katsuda2016} observations.

In this paper, we focus especially on photometric and spectroscopic properties of interacting SNe as a probe of their CSM geometries. \citet{Smith2015} have speculated the photometric and spectroscopic behavior of interacting SNe with a disklike CSM in the context of observations of PTF11iqb \citep[see also \citet{Smith2017} and references therein; e.g.,][]{Andrews2018, Andrews2019}. If an SN has a CSM disk in its vicinity, the majority of the SN ejecta in the polar directions can expand freely as a normal SN. The rest of the SN ejecta in the equatorial direction, however, becomes decelerated by an interaction with the CSM disk. As a result, the opaque SN ejecta in the polar direction covers the CSM interaction region in the equatorial direction, and thus the CSM interaction works as an additional heating source hidden behind the SN ejecta. Therefore, if seen from the polar direction, it will look like a luminous Type II SN, where its spectrum will be like those of normal Type II SNe with P-Cygni lines. Since the photosphere recedes inward, we finally see the interaction directly after a while. After that, we will see it as an interacting SN with a spectrum with narrow lines. \citet{Andrews2018} have suggested this scenario to explain the so-called ``impossible'' SN, iPTF14hls, which shows a very long photospheric phase with normal spectra resembling a traditional Type IIP explosion \citep[][]{Arcavi2017}. In this work we confront the heuristic model of \citet{Smith2015} by calculating light curves of the SNe in their scenario.

There are some hydrodynamics simulations for SNe interacting with a CSM disk \citep[e.g.,][]{McDowell2018, Kurfurst2019, Suzuki2019}.
%Blondin et al. 1996; van Marle et al. 2010; Vlasis et al. 2016; McDowell et al. 2018; Kurfürst & Krtička 2019; Suzuki et al. 2019).
%
However, they did not precisely investigate the radiation properties such as the one qualitatively described above, because, in their simulations, they did not take into account the ionization and recombination, i.e., the fact that the photosphere of Type IIP SNe between $\sim 10$-$\sim 100$ days is formed at a hydrogen recombination front with nearly-fixed radius.
In this work, we analytically calculate light curves for SNe interacting with a CSM disk viewed from the polar direction, and examine the effects of CSM geometry on their light curves. We do not discuss their spectroscopic properties in detail, just assuming that their spectra begin to show narrow lines originated from the interaction once the shock emerges from the photosphere.
%From the results, we propose a new method to investigate the CSM geometry around interacting SNe, through early photometric and spectroscopic observations. 
%
This paper is structured as follows: In Section~2, we describe our light curve model for an interacting SN with a CSM disk. In Section~3, we provide the expected light curves for SNe with various mass and distribution of CSM, and discuss the dependence of the light curve shapes on the parameters. We conclude the paper in Section~4.

%%%%%%%%%%%%%%%%% Light curve model for interacting SNe with aspherical CSM %%%%%%%%%%%%%%%%%%
\section{Light curve model for an interacting SN with a CSM disk}
In this section, we describe our light curve model for an interacting SN with a CSM disk. Figure 1 shows a schematic picture of our model. In the directions toward the disk (Region~1), the SN ejecta interact with the CSM disk, while the SN ejecta expand freely in the polar directions (Region~2). Here we assume, for simplicity, that the motion of the gas fluid is limited in the radial direction, and also that the motion of the SN ejecta is not affected by radiation newly generated by the interaction. Hereafter in this paper, we fix the direction of an observer as the z-axis direction. We also ignore the effects of the disk geometry on the light curve, which can be justified only when the opening angle of the disk is small. In reality, as \citet{Suzuki2019} pointed out, the viewing angle effects and the geometrical effects are important for quantitatively discussing relations between the light curve properties and the total mass/distribution of the CSM. However, the viewing angle effects are beyond a scope of this paper. As a future work, we would like to investigate these points by conducting multi-dimensional radiation hydrodynamics simulations, taking into account the ionization and recombination of hydrogen. Here, we demonstrate the importance of early phase observations of interacting SNe for investigating their CSM geometries with qualitative discussions using a simplified model.

\begin{figure}
	\includegraphics[width=\columnwidth]{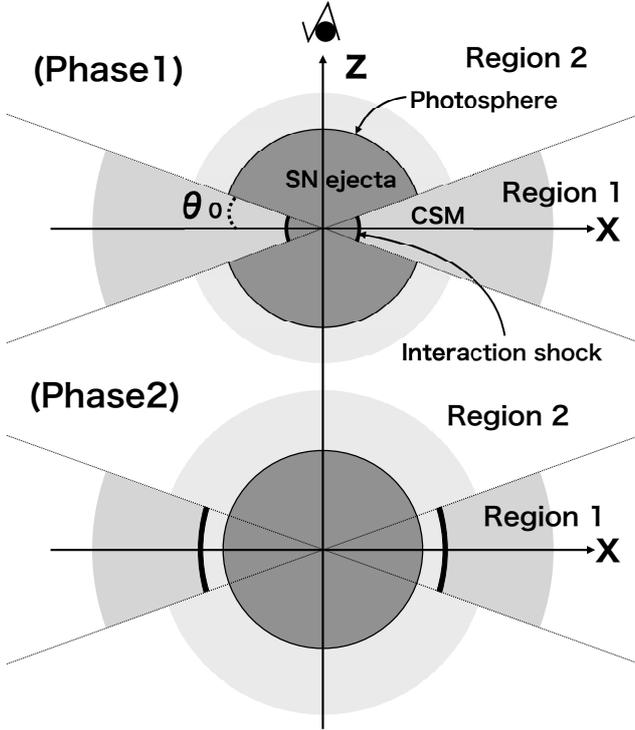}
    \caption{Schematic picture of our model for an SN interacting with a CSM disk in Phase~1 and Phase~2. The observer is assumed to be always in the z-axis direction. The thick black line in Region 1 is the location of the interacting shock. Inside and outside the shock, there are the unshocked ejecta and the unshocked CSM, respectively. The thick black line in Region 2 is the location of the photosphere, which divides the SN ejecta into the optically thick (dark grey) and thin (light grey) regions.}
\end{figure}

%Hydrodynamics

In Region~2, the SN ejecta freely expand as a normal SN. We adopt a simple double power-law distribution for the density of the homologously expanding SN ejecta, which is derived from hydrodynamics simulations \citep[e.g.,][see \S~2.1.1]{Matzner1999}. The SN ejecta shine as a normal Type II SN through photon diffusion processes of the thermal energy, while expanding freely into interstellar space. We determine the SN optical properties in Region~2 using the one-zone analytic model of homologously expanding gas with radiative diffusion by \citet[][see \S~2.1.1]{Kasen2009}. In Region~1, the SN ejecta interact with the CSM disk, converting its kinetic energy into internal energy. We determine the properties of the interaction shock in Region~1 by using a simple self-similar solution of an interaction shock (\S~2.2). The amount of radiation from the interaction shock is determined by a simple model assuming the generated energy leaks out of the expanding shocked shell through photon diffusion processes (\S~2.2). When the shell is located inside the photosphere of the SN ejecta, the radiation from the shock heats the SN ejecta from the inside of the ejecta. This additional heating of the SN ejecta is taken into account by assuming the injected photons are diffusing out through the expanding ejecta (\S~2.3).

%%%%%%%%%%%%%%%%% SN and CSM properties %%%%%%%%%%%%%%%%%%
\subsection{SN and CSM properties}

\subsubsection{SN properties}
We consider an explosion of a red supergiant, which leads to a normal Type IIP SN if there is no CSM. The values of the SN optical properties (luminosity: $L_{\rm{sn}}$, light curve duration: $t_{\rm{sn}}$, and photosphere radius: $R_{\rm{ph}}$) are determined from values of the basic SN ejecta properties (explosion energy: $E_{\rm{sn}}$, ejecta mass: $M_{\rm{ej}}$, pre-SN radius: $R_{\rm{p}}$, and $^{56}$Ni mass: $M_{\rm{Ni}}$) based on the one-zone model of homologously expanding gas with radiative diffusion by \citet{Kasen2009}.

Once the inner part of the progenitor star collapses into a proto-neutron star, a shock wave is created by the core-bounce of the falling material. The shock wave heated by neutrinos from the proto-neutron star proceeds to the surface of the star while heating the ejecta. When it reaches to the surface, the released energy by the shock wave ($E_{\rm{sn}}$) is equally divided into internal energy ($E_{\rm{int}}$) and kinetic energy of the ejecta ($E_{\rm{kin}}$): $E_{\rm{sn}} \sim E_{\rm{int}} \sim E_{\rm{kin}} \sim 10^{51}$ erg. Since a large part of the ejecta is optically thick to electron scatterings, the adiabatic condition can be roughly assumed for the thermal evolution of the ejecta. Thus, we obtain, from the first law of thermodynamics, the internal energy of the expanding ejecta at given time, $t$, taking into account the contribution from radioactive energy input of $^{56}$Ni and $^{56}$Co (see Eq.~9 in \citet{Kasen2009}):
\begin{eqnarray}
\label{eq:eq1}
E_{\rm{int}} (t) &=& E_{\rm{sn}} \left( \frac{R(t)}{R_{\rm{p}}} \right)^{-1} + E_{\rm{Ni}}\frac{t_{\rm{Ni}}}{t} + E_{\rm{Co}}\frac{t_{\rm{Co}}}{t},\\
&=& E_{\rm{sn}} \frac{t_{\rm{e}}}{t} + E_{\rm{Ni}}\frac{t_{\rm{Ni}}}{t} + E_{\rm{Co}}\frac{t_{\rm{Co}}}{t},\\
&=& E_{\rm{sn}} \left( \frac{t_{\rm{e}}}{t} \right) f_{\rm{rad}},
\end{eqnarray}
where
\begin{equation}
f_{\rm{rad}} = 1 + \frac{v_{\rm{sn}}}{E_{\rm{sn}}R_{\rm{p}}}(E_{\rm{Ni}} t_{\rm{Ni}}+E_{\rm{Co}} t_{\rm{Co}}).    
\end{equation}
Here, $R(t)$ is the radius of the expanding gas sphere at time of $t$. Since the ejecta expand roughly homologously, $R(t)=v_{\rm{sn}}t$. The explosion time $t_{\rm{e}}$ is set as $t_{\rm{e}}=R_{\rm{p}}/v_{\rm{sn}}$. $E_{\rm{Ni}}\simeq0.6\times10^{50}(M_{\rm{Ni}}/M_{\odot})$ erg and $E_{\rm{Co}}\simeq1.2\times10^{50}(M_{\rm{Ni}}/M_{\odot})$ erg are the total energy by $^{56}$Ni and $^{56}$Co decay, and $t_{\rm{Ni}}\simeq8.8$ days and $t_{\rm{Co}}\simeq113$ days are the decay time.

The velocity of the ejecta is assumed to be (see Eq.~1 in \citet{Kasen2009}):
\begin{equation}
\label{eq:eq2}
v_{\rm{sn}} = \left( \frac{2E_{\rm{sn}}}{M_{\rm{ej}}} \right)^{1/2}.
\end{equation}

The typical luminosity of the SN ($L_{\rm{sn}}$) is assumed as follows (see Eq.~2 in \citet{Kasen2009}):
\begin{equation}
\label{eq:eq4}
L_{\rm{sn}} = \frac{E_{\rm{int}}(t_{\rm{sn}})}{t_{\rm{sn}}}=\frac{E_{\rm{sn}} t_{\rm{e}}f_{\rm{rad}}}{t_{\rm{sn}}^2}=\frac{E_{\rm{sn}} f_{\rm{rad}} R_{\rm{p}}}{t_{\rm{sn}}^2 v_{\rm{sn}}}.
\end{equation}
Here, the timescale of the light curve, $t_{\rm{sn}}$, is assumed to be the effective diffusion time. Using the mean-free path of photons, $\lambda_{\rm{p}} = (\kappa_{\rm{es}} \rho)^{-1}$, and the optical depth of the ejecta, $\tau = R_{\rm{ph}}/\lambda_{\rm{p}}$, the typical SN duration is expressed as follows (see Eq.~3 in \citet{Kasen2009}):
\begin{equation}
\label{eq:eq5}
t_{\rm{sn}} = \frac{\tau R_{\rm{ph}}}{c} = \frac{R_{\rm{ph}}^2 \kappa_{\rm{es}} \rho}{c},
\end{equation}
where $R_{\rm{ph}}$ is a typical radius of the photosphere in the SN ejecta, $\rho$ ($\simeq M_{\rm{ej}}/v_{\rm{sn}}^{3} t_{\rm{sn}}^{3}$) is the typical density of the ejecta, $\kappa_{\rm{es}}$ is the mass scattering coefficient for the ionized gas. Throughout the paper, we use, for the ionized SN and CSM gas, the value of $\kappa_{\rm{es}} = 0.34$ cm$^{2}$ g$^{-1}$ as opacity in the fully ionized gas composed of hydrogen and helium. The radius of the photosphere is determined mainly by the ionization condition of hydrogen in the ejecta. Thus, the photosphere radius is determined so that the SN luminosity corresponds to the luminosity of a blackbody sphere with a radius $R_{\rm{ph}}$ as follows (see Eq.~7 in \citet{Kasen2009}):
\begin{equation}
\label{eq:eq6}
R_{\rm{ph}} = \left( \frac{L_{\rm{sn}}}{4 \pi \sigma_{\rm{SB}} T_{\rm{I}}^4} \right)^{1/2},
\end{equation}
where, $\sigma_{\rm{SB}}$ is the Stefan-Boltzmann constant and $T_{\rm{I}}$ is the ionization temperature of hydrogen in the SN ejecta. Throughout this paper, we adopt the value, $T_{\rm{I}} = 6000$ K \citep[e.g.,][]{Kasen2009}. Using this expression of $R_{\rm{ph}}$, Eq.~\eqref{eq:eq5} is transformed as
\begin{equation}
\label{eq:eq7}  
  t_{\rm{sn}} = \left( \frac{L_{\rm{sn}}}{4 \pi \sigma_{\rm{SB}} T_{\rm{I}}^4} \right) \frac{\kappa_{\rm{es}}}{c} \frac{M_{\rm{ej}}}{v_{\rm{sn}}^3 t_{\rm{sn}}^3}.
\end{equation}

Now we have three equations for three values $t_{\rm{sn}}$, $L_{\rm{sn}}$ and $R_{\rm{ph}}$: Eq.~\eqref{eq:eq4},\eqref{eq:eq6} and \eqref{eq:eq7}, and an equation for $v_{\rm{sn}}$: Eq.~\eqref{eq:eq2}. Therefore, we can derive the following expressions on these values.
\begin{eqnarray}
\label{eq:eq8}
  t_{\rm{sn}} &=& \left( \frac{\kappa_{\rm{es}}}{16 \pi c \sigma_{\rm{SB}} T_{\rm{I}}^{4}} \right)^{1/6} E_{\rm{sn}}^{-1/6} M_{\rm{ej}}^{1/2} R_{\rm{p}}^{1/6} f_{\rm{rad}}^{1/6},\\
  L_{\rm{sn}} &=& \frac{1}{\sqrt{2}} \left( \frac{16 \pi c \sigma_{\rm{SB}} T_{\rm{I}}^{4}}{\kappa_{\rm{es}}} \right)^{1/3} E_{\rm{sn}}^{5/6} M_{\rm{ej}}^{-1/2} R_{\rm{p}}^{2/3} f_{\rm{rad}}^{2/3}, \\
  \label{eq:r_ph}
  R_{\rm{ph}} &=& \left( \frac{c}{8 \sqrt{2} \pi^{2} \sigma_{\rm{SB}}^{2} \kappa_{\rm{es}} T_{\rm{I}}^{8}} \right)^{1/6} E_{\rm{sn}}^{5/12} M_{\rm{ej}}^{-1/4} R_{\rm{p}}^{1/3} f_{\rm{rad}}^{1/3}.
\end{eqnarray}
Through this paper, we adopt the following values: $E_{\rm{sn}} = 1 \times 10^{51}$ erg, $M_{\rm{ej}} = 10$ M$_{\odot}$ and $R_{\rm{p}} = 500$ R$_{\odot}$, which are typical values for Type IIP SNe \citep[e.g.,][]{Hamuy2003,Smartt2009}. We also use $M_{\rm{Ni}}= 0.032$ M$_{\odot}$, which is the typical $^{56}$Ni mass derived by \citet{Anderson2019} for Type II SNe. The calculated values for the SN optical properties are as follows: $t_{\rm{sn}} \simeq 118.33$ days, $L_{\rm{sn}} \simeq 1.42 \times 10^{42}$ erg s$^{-1}$, $R_{\rm{ph}} \simeq 1.24 \times 10^{15}$ cm.

It is also noted that the above one-zone SN model cannot be applied to the very early phase when the ejecta have not reached to the radius of $R_{\rm{ph}}$ determined within this formalism. To evaluate the duration of this forbidden phase, we consider the timing when the real photosphere in the ejecta reaches to $R_{\rm{ph}}$, which can be calculated as 23.3 days after explosion. Hereafter in this paper, we do not discuss this very early phase. Here, we use the following density profile of the SN ejecta beyond the above one-zone model and assume the real photosphere in the ejecta as a radius where the optical depth to an observer becomes unity. The density profile is assumed to follow the double power-law distribution \citep[e.g.,][]{Moriya2013}, which is derived from hydrodynamics simulations \citep[e.g.,][]{Matzner1999}:
\begin{eqnarray}
\label{eq:rho_ej}
\widetilde{\rho}_{\rm{ej}} (v_{\rm{ej}},t) = 
\left\{
    \begin{array}{l}
      \frac{1}{4 \pi (n-\delta)} \frac{[2(5-\delta)(n-5) E_{\rm{ej}}]^{\frac{n-3}{2}}}{[(3-\delta)(n-3) M_{\rm{ej}}]^{\frac{n-5}{2}}} t^{-3} v_{\rm{ej}}^{-n} \;\;\;\; (v_{\rm{ej}} > v_{t})\\
      \frac{1}{4 \pi (n-\delta)} \frac{[2(5-\delta)(n-5) E_{\rm{ej}}]^{\frac{\delta-3}{2}}}{[(3-\delta)(n-3) M_{\rm{ej}}]^{\frac{\delta-5}{2}}} t^{-3} v_{\rm{ej}}^{-\delta} \;\;\;\; (v_{\rm{ej}} < v_{t})
    \end{array}
  ,\right.
\end{eqnarray}
where $v_{\rm{ej}} (r,t)$ ($= r/t$) is the SN ejecta velocity at radius, $r$, and time, $t$, and 
\begin{equation}
    v_{t} = \left[ \frac{2(5-\delta)(n-5)E_{\rm{ej}}}{(3-\delta)(n-3)M_{\rm{ej}}} \right]^{\frac{1}{2}}.
\end{equation}
Hereafter, $\rho_{\rm{ej}} (r,t)$ is given by $\widetilde{\rho}_{\rm{ej}} (v_{\rm{ej}},t)$ but with the independent variable $v_{\rm{ej}}$ replaced by $r$. We assume that $n=12$ and $\delta = 1$, which are typical values for a red supergiant progenitor \citep[e.g.,][]{Matzner1999}. For $n=12$ and $\delta = 1$, $v_{t}=3.95 \times 10^{3}( E_{\rm{ej}}/1 \times 10^{51} \;\rm{erg} )^{0.5} ( M_{\rm{ej}}/10 M_{\odot})^{-0.5}$ km\;s$^{-1}$. In fact, the above expression of the density profile can apply to the SN ejecta only after reaching to homologous expansion. However, we do not investigate the deviation from the homologous expansion as well as the other values of $n$ and $\delta$ in this paper.

When we consider the CSM interaction (\S~2.2), we use the above density profile of the SN ejecta (Eq.~\eqref{eq:rho_ej}) beyond the above one-zone model.

\subsubsection{CSM properties}
As an initial distribution of the CSM, we adopt the disk structure that has a small value of half-opening angle,  $\theta_{0}$ (see Fig.~1). We use $\theta_{0}=\arcsin{0.1} \sim 5.7^{\circ}$ as a reference value, and also investigate the values of 5, 15, 25, 35 and 45$^{\circ}$ for revealing the effects of $\theta_{0}$ on the light curves. We assume the radial distribution of the CSM as $\rho_{\rm{csm}}(r) = (\dot{M}_{\rm{csm}} / 4\pi v_{\rm{csm}}) r^{-2} = D r^{-2}$ for simplicity, which is expected for the steady mass loss from a progenitor system with constant mass-loss rate of $\dot{M}_{\rm{csm}}$ and wind velocity of $v_{\rm{csm}}$. The velocity of the CSM is set as $v_{\rm{csm}} = 100$ km s$^{-1}$. We investigate the values of $10^{-4}$ to $10^{-1}$ M$_{\odot}$ yr$^{-1}$ for $\dot{M}_{\rm{csm}}$. The inner radius of the disk is assumed to be $R_{\rm{p}}$. In the outer region, the CSM is assumed to be distributed infinitely.

%%%%%%%%%%%%%%%%% CSM interaction %%%%%%%%%%%%%%%%%%
\subsection{CSM interaction}
We determine the luminosity from the interaction shock between the ejecta and the CSM disk ($L_{\rm{disk}}$) using a simple analytical model. This is a modified version of the model by \citet{Moriya2013} including the optical depth effects within the shocked shell on the luminosity. The evolution of the shocked shell in the interaction can be evaluated by the conservation of momentum, assuming the thickness of the shocked shell is much smaller than its radius (see Eq.~1 and 2 in \citet{Moriya2013}).
\begin{eqnarray}
\label{eq:eom}
    M_{\rm{sh}} (t) \frac{dv_{\rm{sh}}(t)}{dt} &=& 4 \pi r_{\rm{sh}}^{2}(t) \Bigl[ \rho_{\rm{ej}} (r_{\rm{sh}}(t),t) \left( v_{\rm{ej}}(r_{\rm{sh}}(t),t)-v_{\rm{sh}}(t) \right)^{2} \nonumber\\
    && - \rho_{\rm{csm}} (r_{\rm{sh}}(t)) \left( v_{\rm{sh}}(t)-v_{\rm{csm}} \right)^{2} \Bigr],
\end{eqnarray}
where $M_{\rm{sh}}(t)$ is the total mass of the shocked SN ejecta and CSM at given time, $t$, and $v_{\rm{sh}}(t)$ is the velocity of the shocked shell at given time, $t$. Here $M_{\rm{sh}}(t)$ is expressed as follows (see Eq.~4 in \citet{Moriya2013}):
\begin{equation}
    M_{\rm{sh}} (t) = \int_{R_{\rm{p}}}^{r_{\rm{sh}}(t)} 4 \pi r^{2} \rho_{\rm{csm}} (r) dr + \int_{r_{\rm{sh}}(t)}^{r_{\rm{ej,max}}(t)} 4 \pi r^{2} \rho_{\rm{ej}} (r,t) dr
\end{equation}
where $r_{\rm{ej,max}}(t) = v_{\rm{ej,max}} t$, and $v_{\rm{ej,max}}$ is the original velocity of the outermost layer of the SN ejecta before the interaction. Here, we assume $r_{\rm{sh}}(t) >> R_{p}$ and $r_{\rm{ej,max}}(t) >> r_{\rm{sh}}(t)$ at all times.

From Eq.~\eqref{eq:eom}, we can analytically derive the time-evolution of the shocked shell and the velocity of the shock as follows, until the time $t_{t}$, when the shock reaches to the inner part of the SN ejecta, i.e., $r_{\rm{sh}}(t_{t}) = v_{t} t_{t}$ (see Eq.~17 and 38 in \citet{Moriya2013}):
\begin{equation}
\label{eq:r_sh}
r_{\rm{sh}} (t) = \left[ \frac{2^{8} \cdot 7^{4.5}}{3^{9} \cdot 11} \frac{E_{\rm{ej}}^{4.5}}{M_{\rm{ej}}^{3.5}} \frac{v_{\rm{csm}}}{\dot{M}_{\rm{csm}}} \right]^{0.1} t^{0.9},
\end{equation}
and
\begin{equation}
\label{eq:v_sh}
v_{\rm{sh}} (t) = \frac{3^{2}}{2 \cdot 5} \left[ \frac{2^{8} \cdot 7^{4.5}}{3^{9} \cdot 11} \frac{E_{\rm{ej}}^{4.5}}{M_{\rm{ej}}^{3.5}} \frac{v_{\rm{csm}}}{\dot{M}_{\rm{csm}}} \right]^{0.1} t^{-0.1}.
\end{equation}
Here,
\begin{eqnarray}
\label{eq:t_t}
    t_{t} = \frac{3}{2^{2} \cdot 7^{0.5} \cdot 11} \frac{M_{\rm{ej}}^{1.5}}{E_{\rm{ej}}^{0.5}} \frac{v_{\rm{csm}}}{\dot{M}_{\rm{csm}}} \simeq 4.2 \times 10^{3} \left( \frac{M_{\rm{ej}}}{10 M_{\odot}} \right)^{1.5} \nonumber\\
    \left( \frac{E_{\rm{ej}}}{1 \times 10^{51} \rm{erg}} \right)^{-0.5} \left( \frac{v_{\rm{csm}}}{10^{7} \rm{cm\;s}^{-1}} \right) \left( \frac{\dot{M}_{\rm{csm}}}{10^{-3} M_{\odot}\; \rm{yr}^{-1}} \right)^{-1} \; \rm{days}.
\end{eqnarray}
For the values of $r_{\rm{sh}} (t)$ and $v_{\rm{sh}} (t)$ after $t_{t}$, we numerically solve the equation of motion with the same initial conditions so that the solutions are continuous to the above analytical values.

%\subsubsection{Bolometric light curves from the shock}
First, we consider light curves for interacting SNe with spherical symmetric CSM. In the optical thin limit, a fraction of the generated energy ($\rm{d}E_{\rm{kin}}(t)/\rm{d}t$) can escape freely from the shocked shell as radiation as follows (see Eq.~20 and 21 in \citet{Moriya2013}):
\begin{equation}
    L(t) = \epsilon \frac{\rm{d}E_{\rm{kin}}(t)}{\rm{d}t} = 2 \pi \epsilon \rho_{\rm{csm}} (r_{\rm{sh}}(t)) r_{\rm{sh}}^{2}(t) v_{\rm{sh}}^{3}(t),
\end{equation}
where $\epsilon$ is the conversion efficiency from kinetic energy to radiation, and
\begin{equation}
    \rm{d}E_{\rm{kin}}(t) = 4 \pi r_{\rm{sh}}^{2}(t) \left( \frac{1}{2} \rho_{\rm{csm}} (r_{\rm{sh}}(t)) v_{\rm{sh}}^{2}(t) \right) \rm{d}r.
\end{equation}
Throughout this paper, we assume $\epsilon=0.1$ \citep[][]{Moriya2013}.

Especially when the optical depth in the shocked shell is more than unity, the optical-depth effects become important. Since the generated photons stay in the shocked shell for diffusion time, their energy becomes reduced due to expansion loss and their emergence becomes delayed. After the escape from the shock, the photons stay also in the CSM for a while. Thus, their arrival to the observer becomes more delayed, but their energy does not become reduced in this diffusion processes within the CSM. This is because the photons have already been decoupled from the shocked gas, i.e., they are already after the shock breakout, where the diffusion time is smaller than the dynamical time of the shock. Here, we do not take into account the effect of the delayed arrival by the diffusion processes within the CSM, because the effect is small compared with that in the shock in our situation. We only take into account the optical depth effects in the shocked shell, using a simple model.
We calculate a light curve from the interaction as a superposition of gas shells with photons that are generated at each time (see Appendix~1 for radiation from each gas shell). The generated photons from time $t-\rm{d}t/2$ to $t+\rm{d}t/2$ experience diffusion processes with $\tau_{\rm{diff}}(t)$ as follows:

\begin{equation}
    \tau_{\rm{diff}}(t) = \kappa_{\rm{es}} \frac{M_{\rm{sh}}(t)}{4 \pi r_{\rm{sh}}^{2} (t) \Delta R_{\rm{sh}}(t)} \Delta R_{\rm{sh}}(t) = \frac{\kappa_{\rm{es}} M_{\rm{sh}}(t)}{4 \pi r_{\rm{sh}}^{2} (t)},
\end{equation}
where $\Delta R_{\rm{sh}}(t)$ is the width of the shocked shell at given time, t.

Thus, the light curve from the interaction with spherical CSM is assumed to be expressed as follows (see Appendix~A):
\begin{equation}
    L_{\rm{sphere}} (t) = \int_{0}^{t} \frac{L_{\rm{sh}}(t^{'}) \rm{d}t^{'}}{t_{\rm{diff}}(t^{'})} \exp \left( - \frac{t-t^{'}}{t_{\rm{diff}}(t^{'})} \left( 1 + \frac{t-t^{'}}{2 t_{\rm{dyn}}(t^{'})} \right) \right),
\end{equation}
where
\begin{eqnarray}
t_{\rm{dyn}} (t) &=& \frac{r_{\rm{sh}}(t)}{v_{\rm{sh}}(t)},\\
t_{\rm{diff}} (t) &=& \frac{\tau_{\rm{diff}}(t) r_{\rm{sh}}(t)}{c}.
\end{eqnarray}

In the case of the interaction with the disk CSM, the CSM is confined into the small region (see Fig.~1), where the mass-loss rate for the disk ($\dot{M}_{\rm{disk}}$) is defined under the assumption that there is no mass loss except for the CSM disk. Thus, the density scale ($D$ in $\rho_{\rm{csm}}(r)=D r^{-2}$) of the CSM disk corresponds to that of the spherical CSM with $\dot{M}_{\rm{csm}}=\dot{M}_{\rm{disk}}/\omega_{\rm{disk}}$. Therefore, the bolometric light curves from interacting SNe with the CSM disk are calculated as follows:
\begin{equation}
\label{eq:Ldisk}
    L_{\rm{disk}} (t,\dot{M}_{\rm{disk}}) = \omega_{\rm{disk}} L_{\rm{sphere}} \left(t,\frac{\dot{M}_{\rm{disk}}}{\omega_{\rm{disk}}} \right),
\end{equation}
where,
\begin{equation}
\omega_{\rm{disk}} = \frac{\Omega_{\rm{disk}}}{4 \pi} = \sin{\theta_{0}}.
\end{equation}
Here, $\Omega_{\rm{disk}}$ is the solid angle of the CSM disk covering the SN light. It is noted that the isotropic mass-loss rate $\widetilde{\dot{M}}_{\rm{csm}}=\dot{M}_{\rm{disk}}/\omega_{\rm{disk}}$ is a good indicator to understand the evolution of the interaction shock.

\begin{figure*}
	\includegraphics[width=0.66\columnwidth]{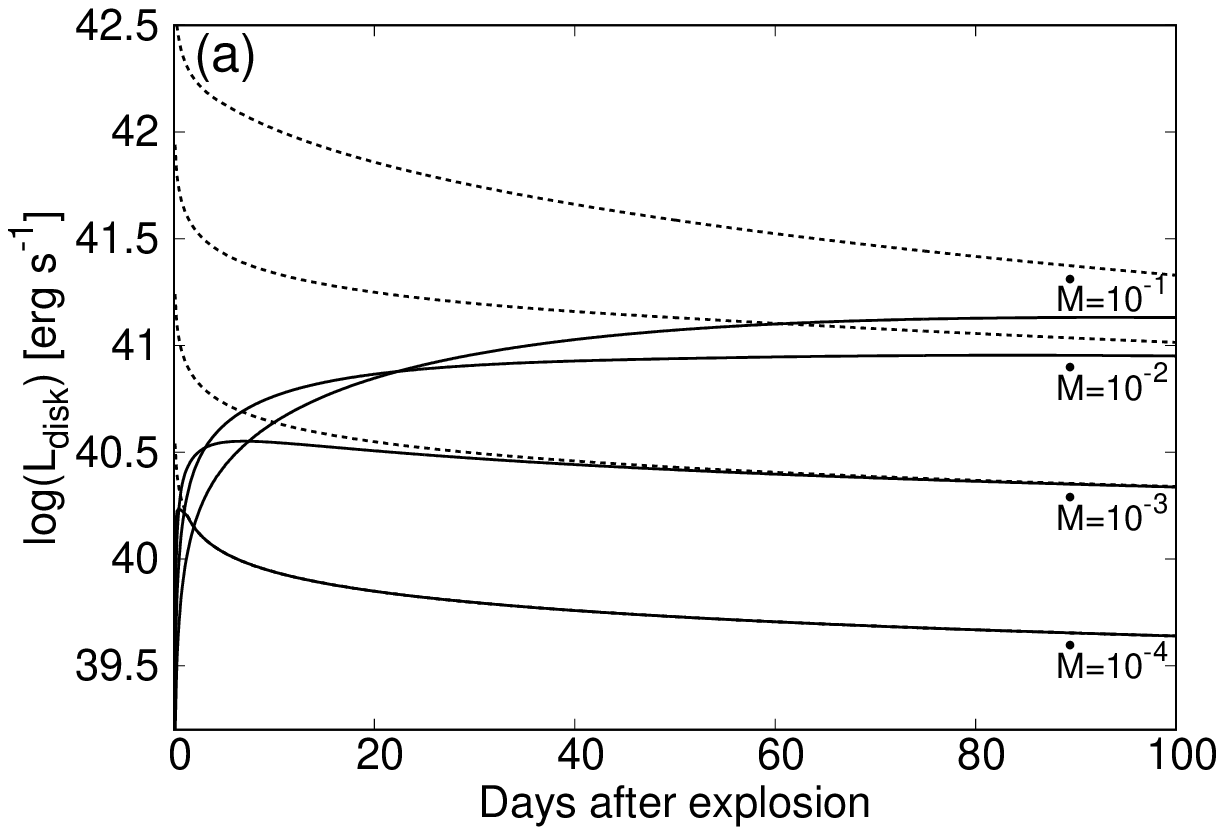}
	\includegraphics[width=0.66\columnwidth]{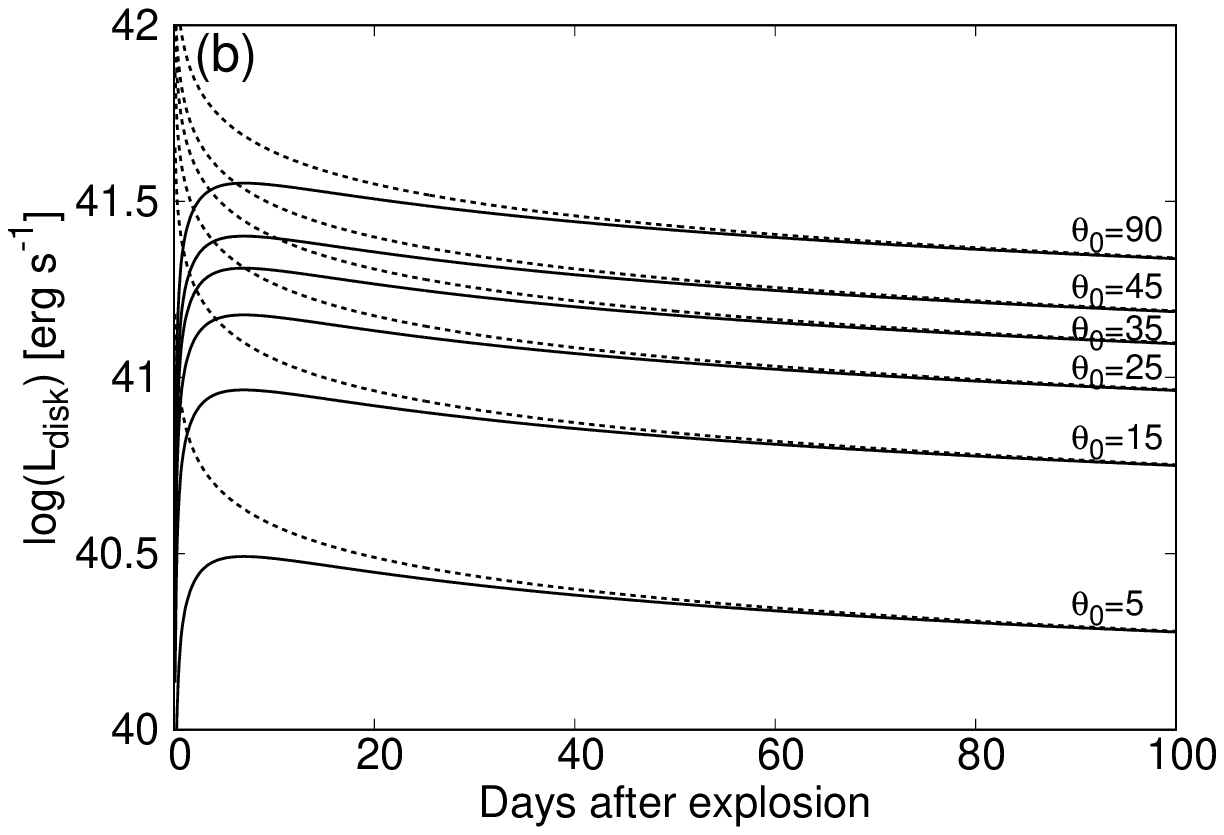}
	\includegraphics[width=0.66\columnwidth]{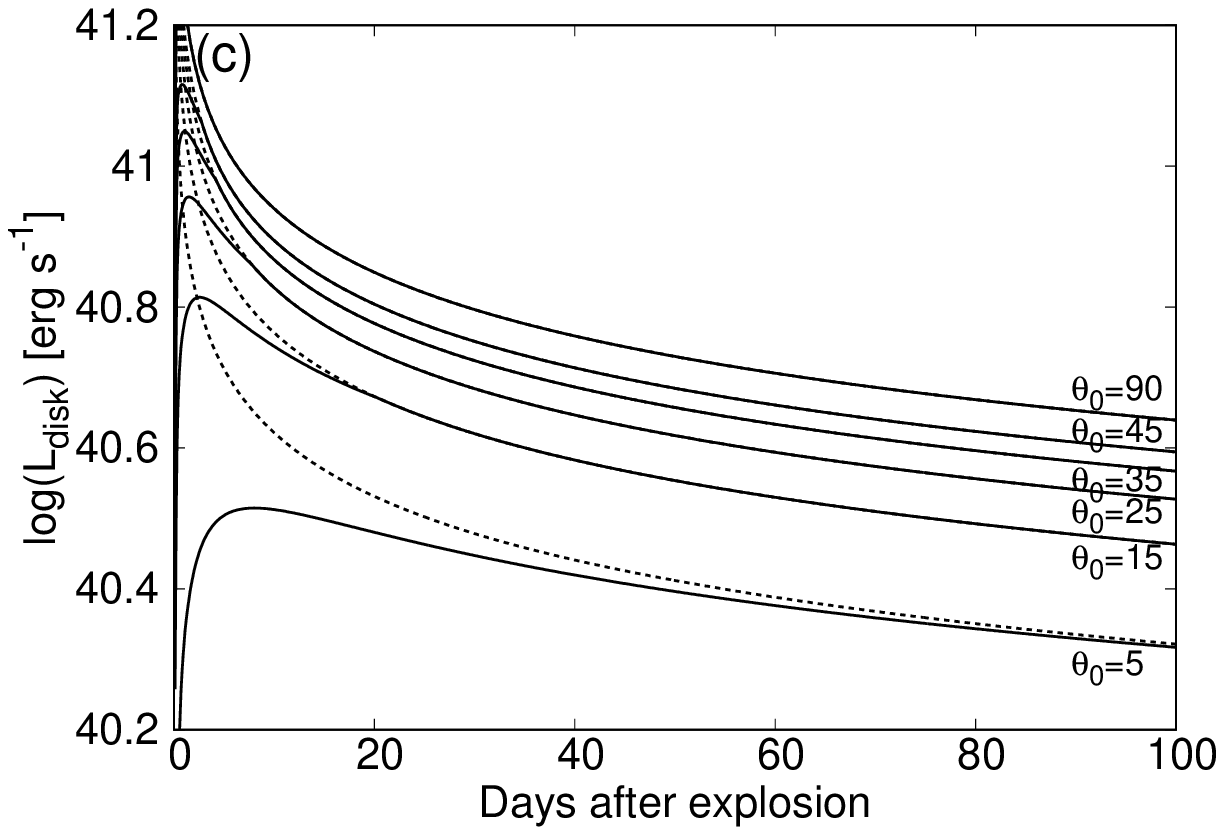}
    \caption{Light curves from the interaction (a) with the CSM disk with $\theta_{0}=\arcsin{0.1} \sim 5.7^{\circ}$ for various values of the mass-loss rate, $\dot{M}_{\rm{disk}}$ ($L_{\rm{disk}}$; solid lines), compared with those in the optical-thin limit (dashed lines); (b) Same as (a), but for the CSM disk with constant $\widetilde{\dot{M}}_{\rm{csm}} = \dot{M}_{\rm{disk}}/\omega_{\rm{disk}} = 10^{-2}$ M$_{\odot}$ yr$^{-1}$ for various values of the opening angles; (c) Same as (a), but for the CSM disk with constant $\dot{M}_{\rm{disk}}= 10^{-3}$ M$_{\odot}$ yr$^{-1}$ for various values of the opening angles.}
\end{figure*}

For the luminosity of the interacting SN with the CSM disk ($L_{\rm{disk}}$), we discuss the dependence of their light curves on the free parameters of the CSM disk ($\dot{M}_{\rm{disk}}$ and $\theta_{0}$). These paramters are connected to the properties of the CSM disk that change the light curve shape from the interaction with the CSM disk: the total mass and density of CSM. The density of the CSM disk determines the evolution of the interaction shock (see Eq.~\eqref{eq:r_sh}), which is related with the luminosity and the above optical-depth effects. On the other hand, the total mass of the CSM disk is important to the luminosity.

Figure~2a shows the input light curves from the interaction with the CSM disk with $\theta_{0}=\arcsin{0.1} \sim 5.7^{\circ}$ for various values of the mass-loss rate, $\dot{M}_{\rm{disk}}$ ($L_{\rm{disk}}$; Eq.~\eqref{eq:Ldisk}). The CSM disks with higher $\dot{M}_{\rm{disk}}$ have higher density and larger total mass of the CSM disk. For higher density of CSM, the generated photons stay longer in the shock through photon diffusion processes, and thus a larger fraction of the generated energy by the interaction is lost due to expansion cooling during the trapping. Especially in the early phase, the light curves differ from those in the optically-thin limit due to the high density. Interestingly SNe with higher $\dot{M}_{\rm{disk}}$ show unusually slow rises in luminosity, while SNe with lower $\dot{M}_{\rm{disk}}$ show rapid rises \citep[e.g.,][]{Moriya2018}. This optical depth effect is important for deriving CSM mass in Type IIn SNe. As a rough indication, we define the duration of the optical-depth effect, $t_{\rm{break}}$, as the time when $t_{\rm{diff}}$ become smaller than $t_{\rm{dyn}}$. This is the same condition for the breakout of the SN shock, which is $\tau_{\rm{diff}} < c/v_{\rm{sh}}$. In the cases in Fig.~2, i.e., the CSM disk with $\theta_{0}=\arcsin{0.1} \sim 5.7^{\circ}$, the values of $t_{\rm{break}}$ are 101.60, 29.71, 2.97 and 0.30 days for $\dot{M}_{\rm{disk}}= 10^{-1}$, $10^{-2}$, $10^{-3}$ and $10^{-4}$ M$_{\odot}$ yr$^{-1}$, respectively. As is seen in Fig.~2a, the timing of $t_{\rm{break}}$ roughly corresponds to the peak of the light curves for each case.

Figure~2b shows the light curves from the interaction with the CSM disk with given $\widetilde{\dot{M}}_{\rm{csm}} = \dot{M}_{\rm{disk}}/\omega_{\rm{fdisk}}$ but for various values of the opening angle of the disk ($L_{\rm{disk}}$; Eq.~\eqref{eq:Ldisk}). This is a comparison between the CSM disks that have the same radial density distribution but different total mass of CSM, i.e., the systems for larger $\theta_{0}$ have larger total mass of CSM. The deviations from the optically-thin limits come from the optical depth effect discussed above. Since the evolution of the interaction shock is the same for all the cases, the light curves have the same optical-depth effects. On the other hand, since the total mass of CSM is different, the light curves for lager $\theta_{0}$ have larger luminosity in proportion to the CSM mass. Therefore, the light curves become identical if they are corrected for the CSM mass.

Figure~2c shows the light curves from the interaction with the CSM disk with given $\dot{M}_{\rm{disk}}$ but for various values of the opening angle of the disk ($L_{\rm{disk}}$; Eq.~\eqref{eq:Ldisk}). This is a comparison between the CSM disks that have the same total mass of CSM but different radial density distribution, i.e., the systems for larger $\theta_{0}$ have lower density. Even though the light curves for different opening angles originate from the interaction with the same total mass of CSM, they show large difference not only in the early phase but also in the later phase. If the CSM is confined to smaller area, i.e., if $\theta_{0}$ is smaller, the optical depth of the interaction shock becomes higher. Therefore, the light curves for smaller $\theta_{0}$ show larger deviations from those in the optically-thin limit. The differences in the late phase, when the shock is already optically thin, are due to the difference of available kinetic energy of the SN ejecta. The available energy is limited by the ejecta kinetic energy within $\theta_{0}$, and thus the shock velocity become slower more rapidly for smaller $\theta_{0}$. Therefore, we obtain lower luminosity for smaller $\theta_{0}$ with the fixed total mass of CSM (i.e., the same value of $\dot{M}_{\rm{disk}}$).

%%%%%%%%%%%%%%%%% Light curve model %%%%%%%%%%%%%%%%%%
\subsection{Light curve model for the interacting SNe with the disk CSM}

\begin{figure}
	\includegraphics[width=\columnwidth]{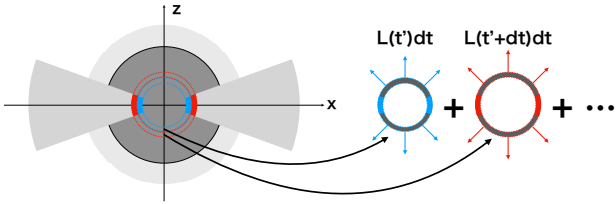}
    \caption{Schematic picture of the procedure for calculating light curves of interacting SNe with the CSM disk.}
\end{figure}

When the interaction shock is below the photosphere in the ejecta (Phase~1), the hidden energy input from the interaction heats the SN ejecta. We consider a situation where the radiation energy injected from Region~1 into Region~2 is diffusing out through the expanding ejecta. We assume that the energy created by the interaction from time $t^{'}-\rm{d}t/2$ to $t^{'}+\rm{d}t/2$ ($L_{\rm{disk}}(t^{'}) \rm{d}t$) is injected into a shell in the ejecta at the radius of the interaction ($r_{\rm{sh}}(t^{'})$; see Fig.~3), and that the injected photons are diffusing out while being affected by the expansion cooling. The ejecta shell have the following diffusion time and dynamical time:
\begin{eqnarray}
    t_{\rm{diff,ej}}(t^{'}) &=& \frac{\tau_{\rm{ej}}(r_{\rm{sh}}(t^{'}),t^{'}) r_{\rm{sh}}(t^{'})}{c}, \\
    t_{\rm{dyn,ej}}(t^{'}) &=& \frac{r_{\rm{sh}}(t^{'})}{v_{\rm{ej}}(r_{\rm{sh}}(t^{'}),t^{'})} = t^{'},
\end{eqnarray}
where 
\begin{equation}
\label{eq:tauej}
    \tau_{\rm{ej}} (r,t) = \int_{r}^{\widetilde{R}_{\rm{ph}}(t)} \kappa_{\rm{es}} \rho_{\rm{ej}} (r^{'},t) \rm{d}r^{'}.
\end{equation}
%Thus, the light curve only by the injected photons during $t^{'}-\rm{d}t/2$ to $t^{'}+\rm{d}t/2$ is calculated as follows:
%\begin{equation}
%    \frac{L_{\rm{disk}}(t^{'}) \rm{d}t^{'}}{t_{\rm{diff,ej}}(t^{'})} \exp \left( - \frac{t-t^{'}}{t_{\rm{diff,ej}}(t^{'})} \left( 1 + \frac{t-t^{'}}{2 t_{\rm{dyn,ej}}(t^{'})} \right) \right).
%\end{equation}
%
Until the interaction region is exposed to an observer, the additional heating from the hidden interaction contributes to the SN light in addition to the original thermal energy ($L_{\rm{sn}}$). Therefore, the light curve from the system can be described as follows (see Appendix~A): 
\begin{equation}
\label{eq:Ltot}
    L_{\rm{tot}} (t) = L_{\rm{sn}}(t) + \int_{0}^{t} \frac{L_{\rm{disk}}(t^{'}) \rm{d}t^{'}}{t_{\rm{diff,ej}}(t^{'})} \exp \left( -\frac{t-t^{'}}{t_{\rm{diff,ej}}(t^{'})} \left( 1 + \frac{t-t^{'}}{2t_{\rm{dyn,ej}}(t^{'})} \right) \right),
\end{equation}
where the luminosity from the original SN thermal energy is given as $L_{\rm{sn}}$ until $t=t_{\rm{sn}}$, and then as zero (see \S 2.1.1).
%\begin{eqnarray}
%L_{\rm{sn}}(t) = \left\{ \begin{array}{ll}
%    \frac{1}{\sqrt{2}} \left( \frac{16 \pi c \sigma_{\rm{SB}} T_{I}^{4}}{\kappa_{\rm{es}}} \right)^{1/3} E_{\rm{sn}}^{5/6} M_{\rm{ej}}^{-1/2} R_{\rm{p}}^{2/3} & (t \leqq t_{\rm{sn}}) \\
%    0 & (t > t_{\rm{sn}})
%  \end{array} \right.
%\end{eqnarray}
Finally, the photosphere radius in the ejecta is calculated as follows:
\begin{equation}
\label{eq:Rph}
    \widetilde{R}_{\rm{ph}} (t) = \left( \frac{L_{\rm{tot}}(t)}{4 \pi \sigma_{\rm{SB}} T_{I}^{4}} \right)^{1/2}.
\end{equation}

For calculating $L_{\rm{tot}}(t)$ (i.e., $\tau_{\rm{ej}}$ in Eq. \eqref{eq:tauej}), we need to know the value of $\widetilde{R}_{\rm{ph}} (t^{'})$ for each time of $t^{'} \leqq t$. At the same time, we need the value of $L_{\rm{tot}}(t)$ to calculate $\widetilde{R}_{\rm{ph}} (t)$. Therefore, we perform interative processes for obtaining the values for $L_{\rm{tot}}(t)$ and $\widetilde{R}_{\rm{ph}} (t)$. First we calculate $L_{\rm{tot}} (t^{'}+\rm{d}t)$ assuming $\widetilde{R}_{\rm{ph}} (t^{'}+\rm{d}t) = \widetilde{R}_{\rm{ph}} (t^{'})$, when the values of $\widetilde{R}_{\rm{ph}} (t)$ are known for $t \leqq t^{'}$. Then the value of $\widetilde{R}_{\rm{ph}} (t^{'}+\rm{d}t)$ is modified from Eq. \eqref{eq:Rph} using the obtained value of $L_{\rm{tot}} (t^{'}+\rm{d}t)$. Through these iterative processes, we obtain the final values for $L_{\rm{tot}}(t)$ and $\widetilde{R}_{\rm{ph}} (t)$.

Once the shock radius ($r_{\rm{sh}} (t)$) exceeds this radius of $\widetilde{R}_{\rm{ph}} (t)$ (Phase~2), the situation is changed. This timing of the dramatic change is assigned as $t_{\rm{hidden}}$ here: $r_{\rm{sh}} (t_{\rm{hidden}}) = \widetilde{R}_{\rm{ph}} (t_{\rm{hidden}})$. After this timing, radiation from the system is a sum of the original SN light, the remnant photons that were injected by the interaction shock and are still in the ejecta, and the direct light from the interaction (Phase~2):
\begin{eqnarray}
\label{eq:Ltot2}
    L_{\rm{tot}}(t) &=& L_{\rm{sn}} (t) \nonumber\\
    &+& \int_{0}^{t_{\rm{hidden}}} \frac{L_{\rm{disk}}(t^{'}) \rm{d}t^{'}}{t_{\rm{diff,ej}}(t^{'})} \exp \left( -\frac{t-t^{'}}{t_{\rm{diff,ej}}(t^{'})} \left( 1 + \frac{t-t^{'}}{2t_{\rm{dyn,ej}}(t^{'})} \right) \right) \nonumber \\
    &+& L_{\rm{disk}}(t).
\end{eqnarray}
During Phase~2, the photosphere radius in the ejecta is calculated using the luminosity from the SN ejecta (the first and second terms of Eq.~\eqref{eq:Ltot2}) using Eq.~\eqref{eq:Rph}.

\section{Results and discussions}

\begin{figure}
	\includegraphics[width=\columnwidth]{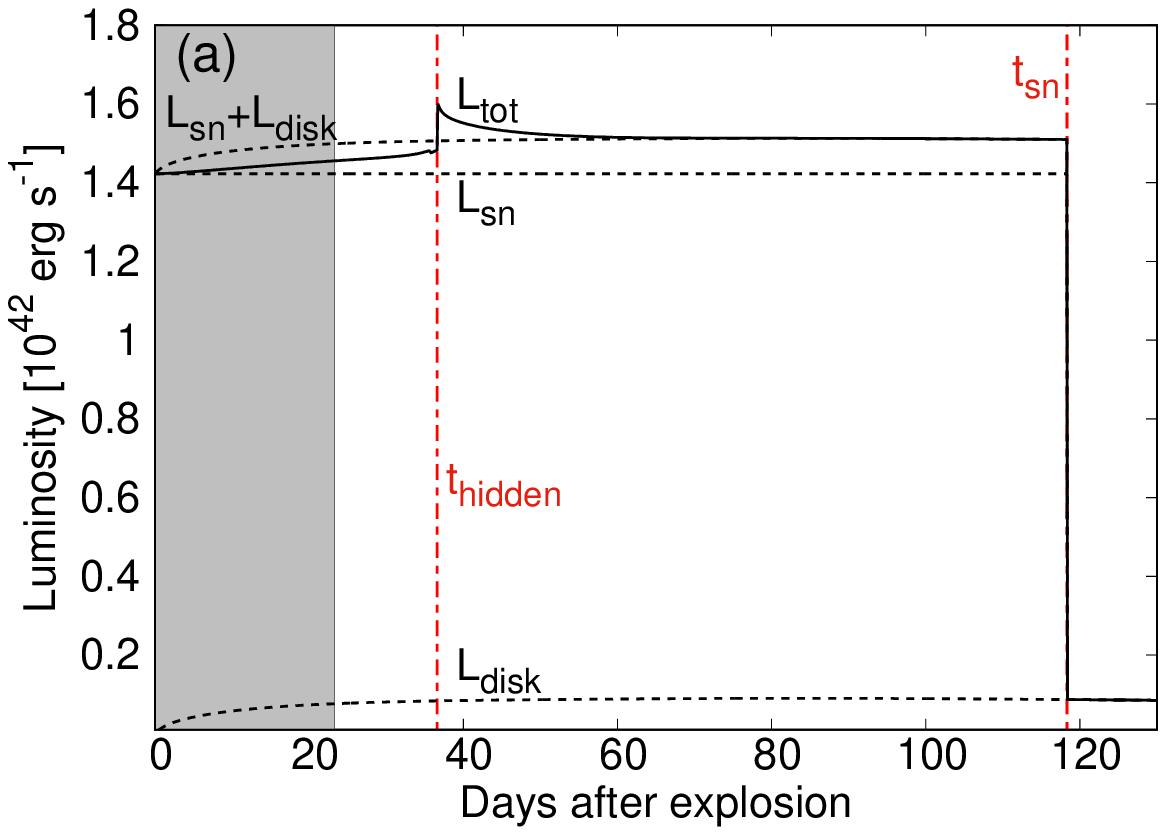}
	\includegraphics[width=\columnwidth]{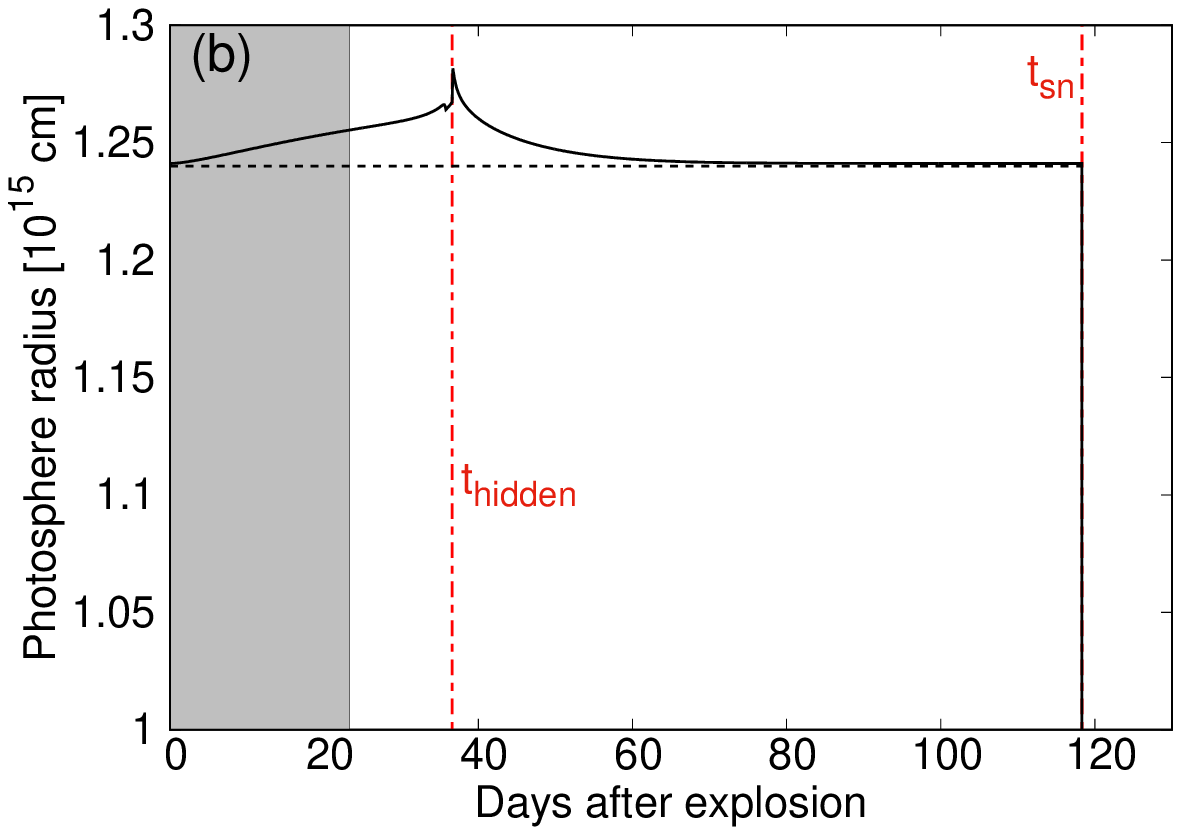}
    \caption{(a) Luminosity from the interacting SN with the CSM disk ($L_{\rm{tot}}$), where $\dot{M}_{\rm{disk}}=10^{-2}$ M$_{\odot}$ yr$^{-1}$ and $\theta_{0}=\arcsin (0.1) \sim 5.7^{\circ}$. The dashed lines show $L_{\rm{sn}}(t)+L_{\rm{disk}}(t)$, $L_{\rm{sn}}(t)$ and $L_{\rm{disk}}(t)$. (b) Time-evolution of the photosphere radius for the same system as Fig. (a). The dashed line shows the radius of $R_{\rm{ph}}$. The dashed-dotted lines show the timing of  $t_{\rm{hidden}}$ and $t_{\rm{sn}}$, which are $19.1$ and $93.0$ days in this case, respectively. The early phase when the present model does not apply is shown by the shaded region. 
    }
\end{figure}

Figure 4a shows a light curve from an interacting SN with the CSM disk with $\dot{M}_{\rm{disk}}=10^{-2}$ M$_{\odot}$ yr$^{-1}$ and $\theta_{0}=\arcsin (0.1) \sim 5.7^{\circ}$. The light curve shape is dramatically changed at a certain time, which corresponds to the time, $t_{\rm{hidden}}$, when the interaction shock in the disk reaches to the radius of the photosphere in the SN ejecta. Before this timing (Phase~1), the energy generated in the interaction shock is injected into the ejecta and reaches to the observer after experiencing the expansion cooling within the ejecta. As is shown in Fig.~4, the luminosity and photosphere radius of the SN in this phase are slightly enhanced due to the hidden energy input, compared to those of the original input SN component. After $t_{\rm{hidden}}$ (Phase~2), the luminosity of the SN becomes a combination of the original SN light, the light from the exposed disk interaction, and the delayed photons generated by the interaction before $t_{\rm{hidden}}$ due to diffusion processes within the ejecta. Thus, a rapid rise in luminosity is observed at this transitional time. At this timing, the spectrum of the SN can be also changed from a spectrum originated from the ejecta (a normal SN spectrum with P-Cygni lines) to that from the shock (an interacting SN spectrum with narrow lines), although it depends on the strength of the interaction shock. Once the thermal energy in the SN ejecta is all lost at $t=t_{\rm{sn}}$ (Phase~3), the luminosity from the system becomes decreased into the level of only the luminosity by the CSM interaction.

\begin{figure}
	\includegraphics[width=\columnwidth]{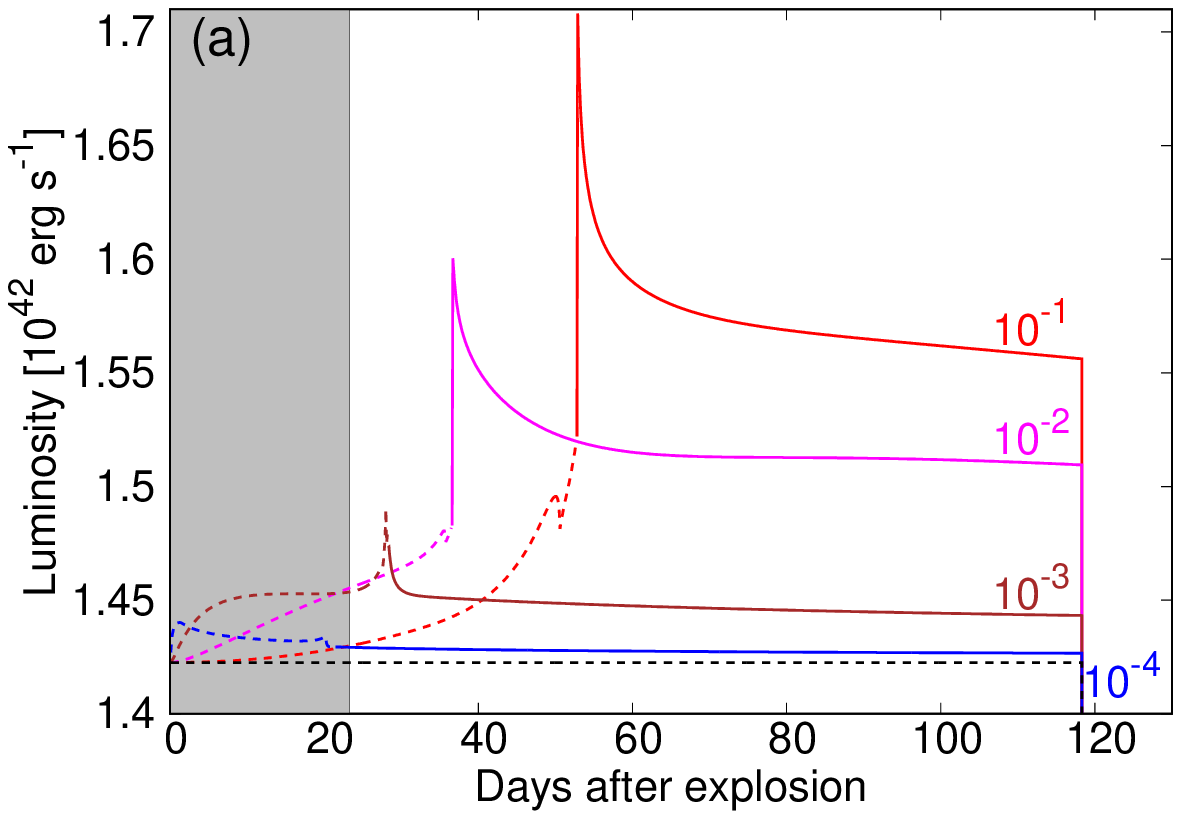}
	\includegraphics[width=\columnwidth]{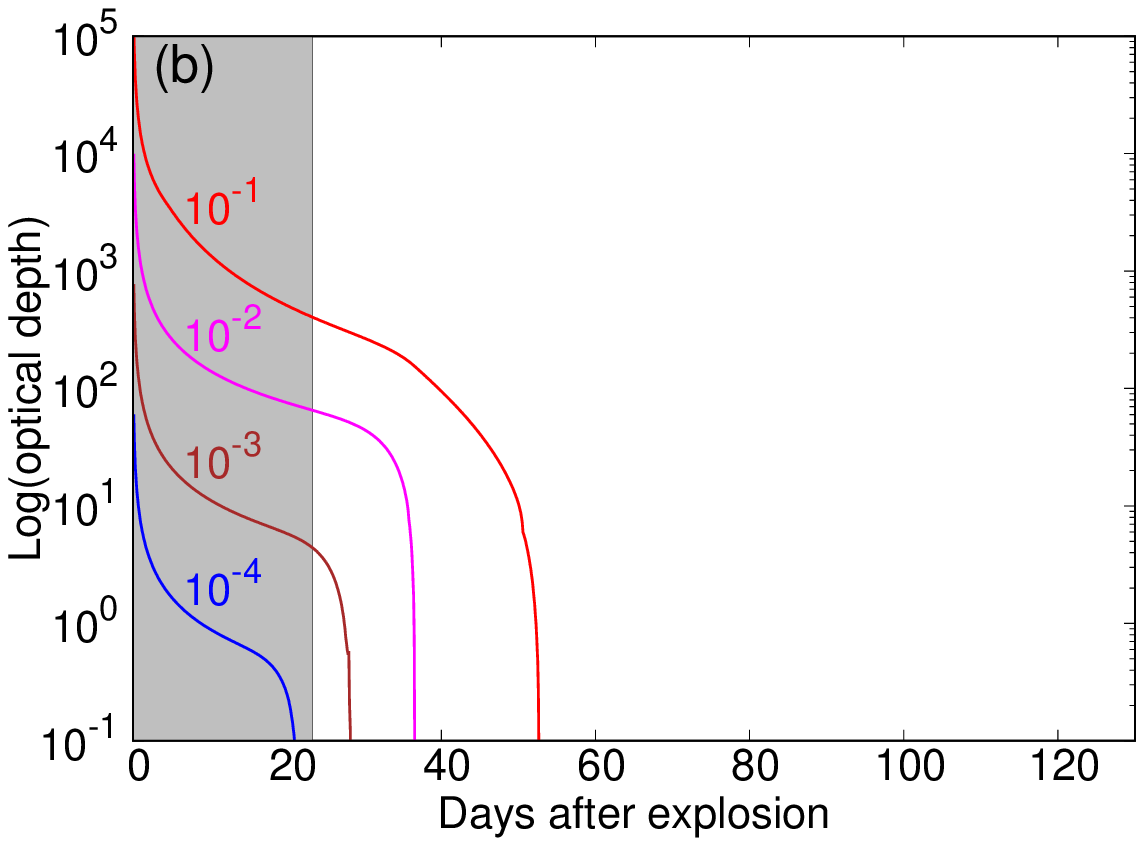}
    \caption{(a) Light curves from the systems with various values of the mass-loss rate of the CSM disk ($\dot{M}_{\rm{disk}}$ in the unit of M$_{\odot}$ yr$^{-1}$), where $\theta_{0}=\arcsin{0.1} \sim 5.7^{\circ}$. The dashed and solid lines express the terms when the system has a spectrum originated from the SN ejecta and from both the SN ejecta and the interaction, respectively. The black dashed line expresses the original SN luminosity ($L_{\rm{sn}}$). (b) Time evolution of the optical depth from $r_{\rm{sh}}$ to $\widetilde{R}_{\rm{ph}}$ in the SN ejecta ($\tau_{\rm{ej}}(r_{\rm{sh}}(t),t)$; Eq.~\eqref{eq:tauej}) for various values of mass-loss rate (in the unit of M$_{\odot}$ yr$^{-1}$). The early phase when the present model does not apply is shown by the shaded region.
    }
\end{figure}

For the interacting SNe with the CSM disk, we discuss the dependence of their light curves on the free parameters of the CSM disk ($\dot{M}_{\rm{disk}}$ and $\theta_{0}$). These correspond to two physical parameters of the CSM disk that change the light curve shape from the SN interaction with the CSM disk: the total mass and density of CSM, as we discussed in \S~2.2 and Fig.~2. The total mass plays a role in determining the luminosity, which also affects the enhancement of the photosphere. On the other hand, the radial density distribution plays a role in determining the evolution of the interaction shock (see Eq.~\eqref{eq:r_sh}), which is related with the luminosity and the optical-depth effects.

Figure 5a shows light curves for the interacting SNe with the CSM disk with given $\theta_{0}$ but for various values of $\dot{M}_{\rm{disk}}$. The CSM disk with higher $\dot{M}_{\rm{disk}}$ has higher density and larger total mass of CSM. Thus, the input luminosity from the CSM disk is basically higher for the system with higher $\dot{M}_{\rm{disk}}$, as discussed in Fig.~2a. The transitional timing, $t_{\rm{hidden}}$, is more delayed for the systems with higher CSM density, mainly because the interaction region is deeper for the systems with higher CSM density as shown in Eq.~\eqref{eq:r_sh} and partly because the enhancement of the photosphere due to the input light from the disk interaction is higher. In addition, the energy lost by the expansion cooling within the ejecta is also higher for the systems with higher CSM density. As shown in Fig.~5b, the optical depth from $r_{\rm{sh}}(t)$ to $\widetilde{R}_{\rm{ph}}$ ($\tau_{\rm{ej}}(r_{\rm{sh}}(t),t)$; Eq.~\eqref{eq:tauej}) is higher for the systems with higher CSM density, because their interaction region is deeper. The optical depth for any cases in Fig.~5b keeps high just before the timing of the shock appearance (i.e., $t_{\rm{hidden}}$), which tells that the expansion cooling is efficient almost all the time when the shock is inside the ejecta. This is because of the steep density structure of the SN ejecta (see Eq.~\eqref{eq:rho_ej}). The luminosity of the SN, soon after $t_{\rm{hidden}}$, becomes the sum of the original SN light and the light from the exposed disk interaction. Thus, the system with higher $\dot{M}_{\rm{disk}}$ is basically brighter as is shown in Fig.~2a.

\begin{figure}
	\includegraphics[width=\columnwidth]{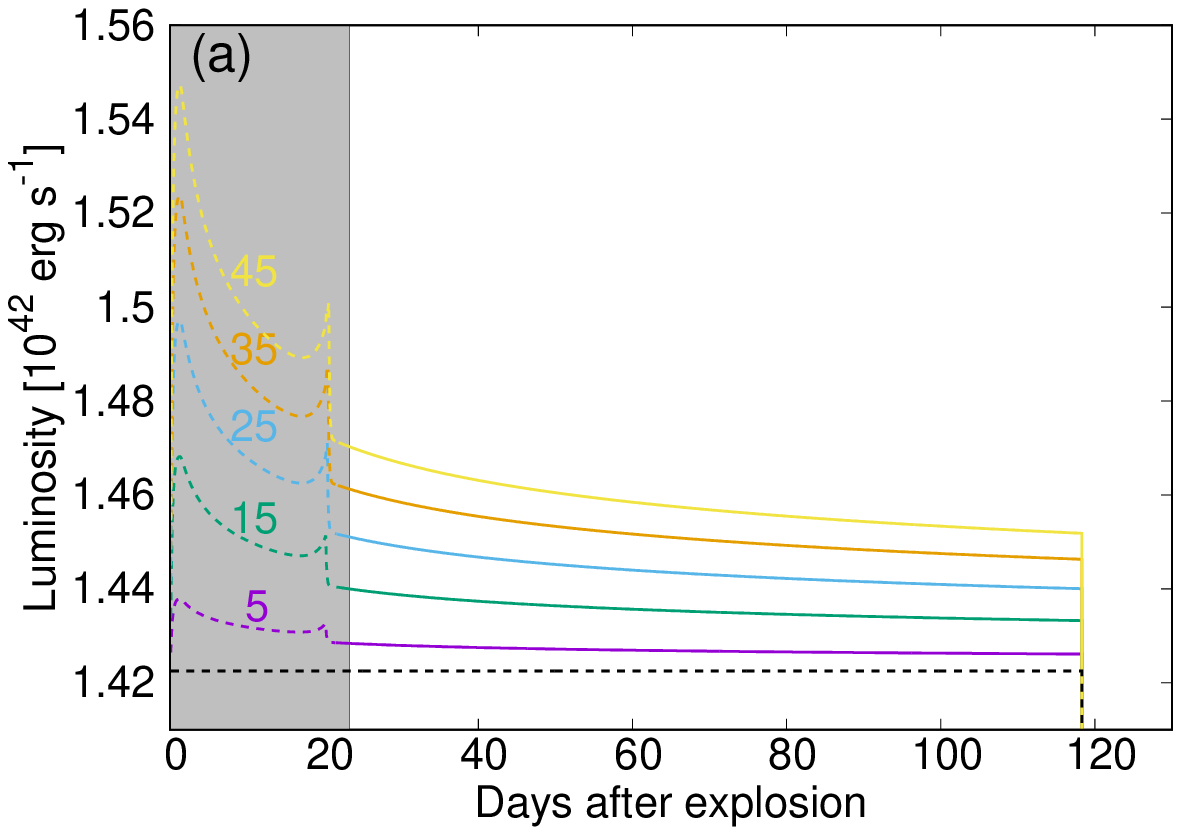}
	\includegraphics[width=\columnwidth]{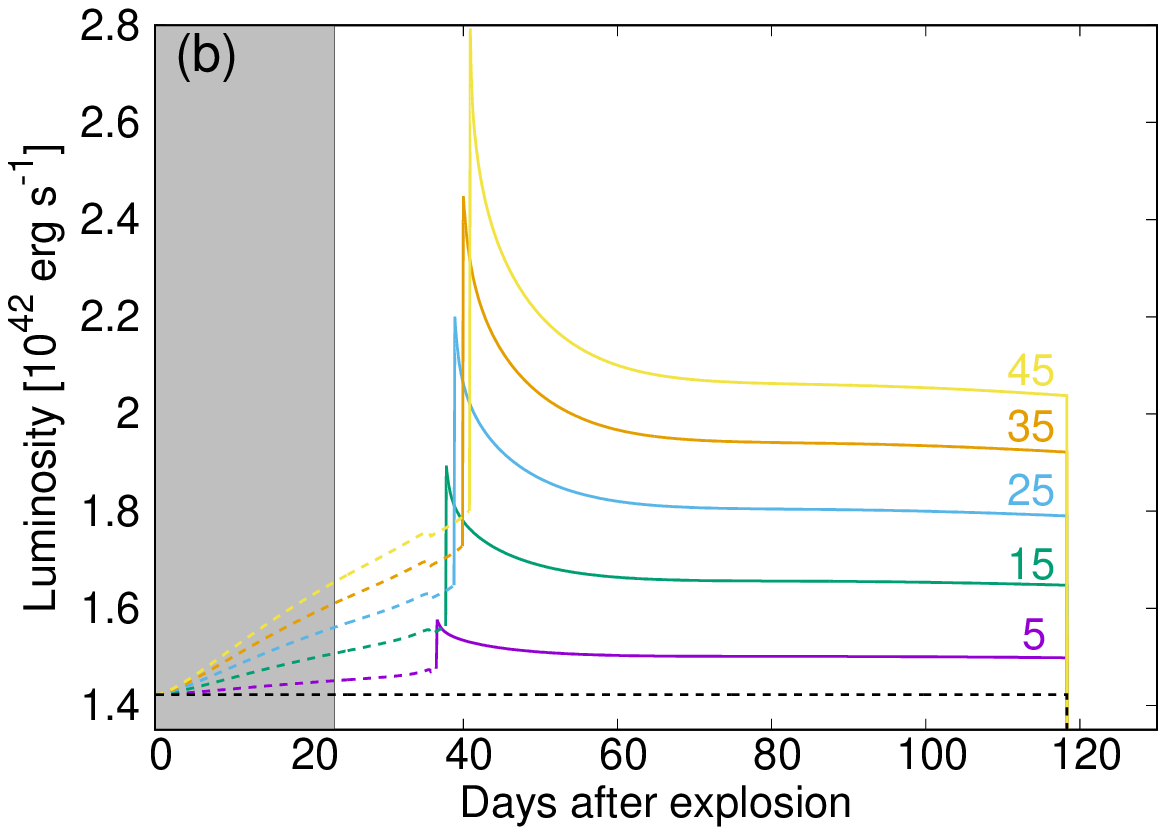}
    \caption{Light curves from the systems with various values of the opening angles of the CSM disk. Here, (a) $\widetilde{\dot{M}}_{\rm{csm}} = \dot{M}_{\rm{disk}}/\omega_{\rm{disk}} = 10^{-3}$ M$_{\odot}$ yr$^{-1}$ and (b) $\widetilde{\dot{M}}_{\rm{csm}} = \dot{M}_{\rm{disk}}/\omega_{\rm{disk}} = 10^{-1}$ M$_{\odot}$ yr$^{-1}$. The dashed and solid lines express the terms when the system has a spectrum originated from the SN ejecta and from both the SN ejecta and the interaction, respectively. The black dashed line expresses the original SN luminosity ($L_{\rm{sn}}$). The early phase when the present model does not apply is shown by the shaded region.
    }
\end{figure}

\begin{figure}
	\includegraphics[width=\columnwidth]{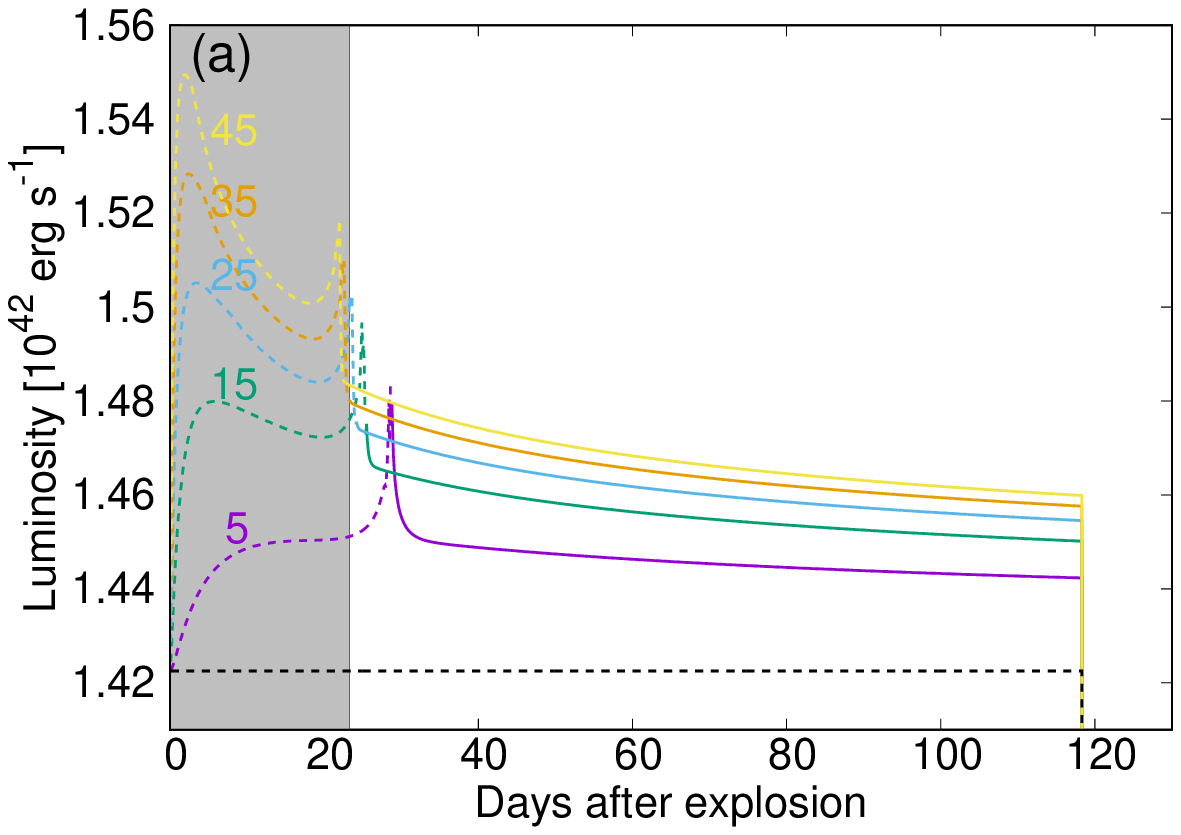}
	\includegraphics[width=\columnwidth]{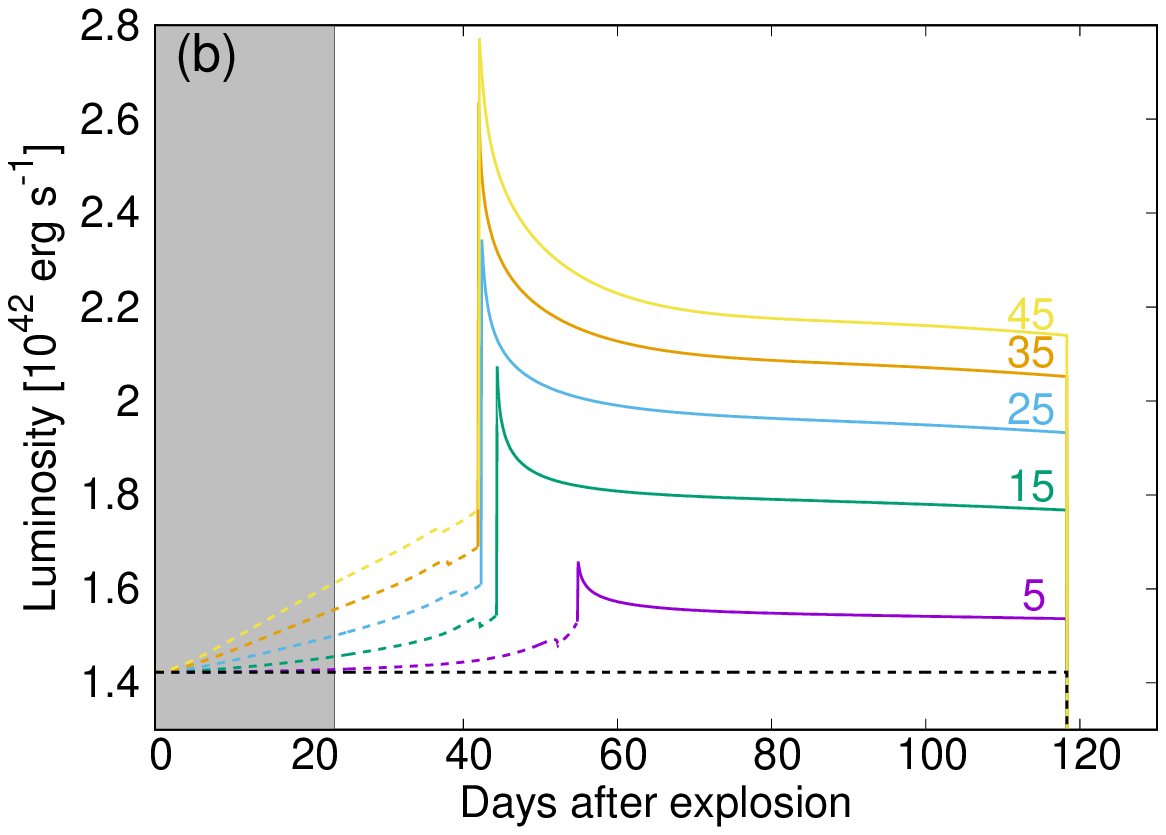}
    \caption{Same as Fig.~6, but for the disk with (a) $\dot{M}_{\rm{disk}}=10^{-3}$ and (b) $\dot{M}_{\rm{disk}}=10^{-1}$ M$_{\odot}$ yr$^{-1}$.
    }
\end{figure}

Figure 6 shows light curves for the interacting SNe with the CSM disk with given $\widetilde{\dot{M}}_{\rm{csm}} = \dot{M}_{\rm{disk}}/\omega_{\rm{disk}}$ but for various values of $\theta_{0}$. The CSM disk for the system with larger $\theta_{0}$ has larger total mass of CSM, while all the disks has constant CSM density. Thus, the input luminosity from the CSM disk is higher for the system with larger $\theta_{0}$ as discussed in \S~2.2 and Fig.~2b. Since the CSM density is the same value for all the cases, the evolution of the shock is identical. In the case of $\widetilde{\dot{M}}_{\rm{csm}} = \dot{M}_{\rm{disk}}/\omega_{\rm{disk}} = 10^{-3}$ M$_{\odot}$ yr$^{-1}$ (Fig.~6a), the transitional timing, $t_{\rm{hidden}}$, is identical for all the cases, because the enhancement of the photosphere due to the hidden energy input is not efficient. On the other hand, $t_{\rm{hidden}}$ for larger $\theta_{0}$, in the case of $\widetilde{\dot{M}}_{\rm{csm}} = \dot{M}_{\rm{disk}}/\omega_{\rm{disk}} = 10^{-1}$ M$_{\odot}$ yr$^{-1}$ (Fig.~6b), is more delayed due to the enhancement of the photosphere.

Figure 7 shows light curves for the interacting SNe with the CSM disk with given $\dot{M}_{\rm{disk}}$ but for various values of $\theta_{0}$. The CSM disk for the system with larger $\theta_{0}$ has higher density of CSM, while all the disks has constant total mass of CSM. Thus, the input luminosity from the CSM disk is higher for the system with larger $\theta_{0}$ as discussed in \S~2.2 and Fig.~2c. The transitional timing, $t_{\rm{hidden}}$, is more delayed for the systems with smaller $\theta_{0}$, if we can ignore the effects of the enhancement of the photosphere due to the hidden energy input, which is the case of Fig.~7a. This is because the interaction region is deeper for the systems with larger CSM density as shown in Eq.~\eqref{eq:r_sh}. In Fig.~7b, $t_{\rm{hidden}}$ is not always higher for smaller $\theta_{0}$, because the photosphere radius is more enhanced for larger $\theta_{0}$ due to larger hidden energy input. In addition, the energy lost by the expansion cooling within the ejecta is also higher for the systems with larger CSM density, i.e., smaller $\theta_{0}$.

\begin{figure}
	\includegraphics[width=\columnwidth]{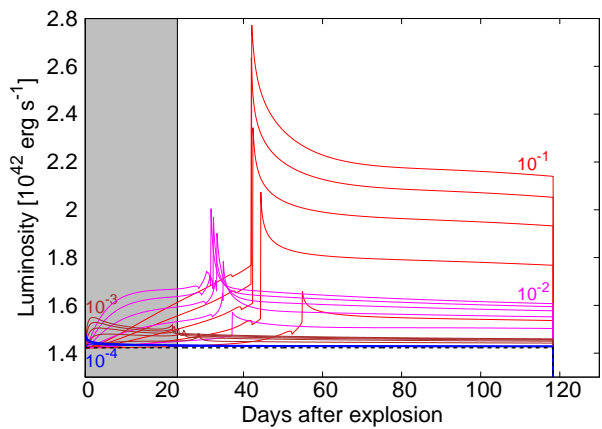}
    \caption{Light curves from the interaction for the systems with various values of mass-loss rate ($\dot{M}_{\rm{disk}}$) and opening angle of the CSM disk ($\theta_{0}$). The red, magenta, brown and blue lines corresponds to the cases with $10^{-1}$, $10^{-2}$, $10^{-3}$ and $10^{-4}$ M$_{\odot}$ yr$^{-1}$, respectively. The cases with the opening angle of $\theta_{0}=5$, 15, 25, 35 and 45$^{\circ}$ are shown (solid lines). The black dashed line expresses the original SN luminosity ($L_{\rm{sn}}$). The early phase when the present model does not apply is shown by the shaded region.
    }
\end{figure}

%\begin{figure}
%	\includegraphics[width=\columnwidth]{fig9.eps}
%    \caption{Luminosity ratio between those at 15 days and at 90 days after explosion for %\end{figure}

Figure 8 shows light curves for the systems with various values of mass-loss rate ($\dot{M}_{\rm{disk}}$) and opening angles ($\theta_{0}$) for the CSM disk. As we have discussed above, $\theta_{0}$, which is related to the density of the CSM disk, determines the degree of the expansion cooling of the hidden energy input in the early phase for given $\dot{M}_{\rm{disk}}$. In addition, the luminosity after $t_{\rm{hidden}}$ depends on the luminosity of the disk interaction, i.e., $\dot{M}_{\rm{disk}}$. Therefore, the ratio of the luminosity before and after $t_{\rm{hidden}}$ can be useful to investigate the opening angle of the disk. It should be noted that we need to conduct more detailed calculations to quantitatively derive the geometry of the CSM disk because we use several critical simplifications in the calculation. One of them is the viewing angle of an observer, which is fixed exactly to the polar direction. Here, we just demonstrate the importance of early phase observations of the interacting SNe for investigating their CSM geometries with qualitative discussions using a simplified model.

%\begin{figure}
%	\includegraphics[width=\columnwidth]{fig9a.eps}
%    \includegraphics[width=\columnwidth]{fig9b.eps}
%    \caption{Light curves from the interacting SNe with the CSM disk with finite outer radii. Here, $\theta_{0}=\arcsin{0.1} \sim 5.7^{\circ}$ and $\dot{M}_{\rm{disk}}=10^{-3}$ (a) and $10^{-2}$ (b) M$_{\odot}$ yr$^{-1}$. The dashed lines express the phases when the SN shows Type II-like spectrum, and the solid lines express the phases when the SN shows Type IIn-like spectrum. The black dashed line expresses the original SN luminosity ($L_{\rm{sn}}$). The early phase when the present model does not apply is shown by the shaded region.
%    }
%\end{figure}

%Figure 9 shows light curves in cases where the disk has a finite outer radius. If the outer radius is small enough, we do not observe the CSM interaction features in spectra. Such SNe show a luminosity bump with normal IIP spectrum with P-Cygni lines, which might explain observations for some Type II SNe with a bump \citep[e.g.,][]{Nakaoka2018}.

%%%%%%
%Duration of hidden energy source
%%%%%%

In the context of interacting SNe with aspherical CSM, some researches have proposed the origin of SNe that show high luminosity with normal spectra of SNe II as the energy input from the hidden CSM interaction \citep[e.g.,][]{Smith2017, Andrews2018, Andrews2019}. In other words, they have proposed the existence of SNe with large $t_{\rm{hidden}}$ in our SN model. Contrary to the earlier explanation, we do not obtain high luminosity by the hidden energy input as we have discussed above. If we have very dense CSM in the vicinity of the SN, the loss of the energy from the shock within the ejecta becomes efficient because the energy is injected in the very inside of the expanding ejecta. We have also demonstrated the difficulty of creating long $t_{\rm{hidden}}$. If there would be no effects on the enhancement of the photosphere radius from the hidden energy input, we can calculate $t_{\rm{hidden}}$ by comparing $r_{\rm{sh}}$ (Eq.~\eqref{eq:r_sh}) with $R_{\rm{ph}}$ (Eq.~\eqref{eq:r_ph}) derived above by the analytical calculations. The calculated values are shown in Table~1. In reality, the hidden energy input slightly extends the photosphere radius in the SN ejecta. The values of $t_{\rm{hidden}}$ from our model, including this effect, are also shown in Table~1. Here, in the cases with $\dot{M}_{\rm{disk}} \lesssim 2.0 \times 10^{-4}$ M$_{\odot}$ yr$^{-1}$, where $t_{\rm{hidden}} < 23.3$ days (the early phase when the present model does not apply), the real values of $t_{\rm{hidden}}$ should be smaller than those in Table~1. This is because the real photosphere is below $R_{\rm{ph}}$ during this forbidden phase (see the discussions in 2.1.1). If we increase the CSM density further, $t_{\rm{hidden}}$ will be increased but, at the same time, $t_{\rm{break}}$ will also be increased (see the above discussion on Fig.~2a in \S 2.2). Eventually, $t_{\rm{break}}$ becomes bigger than $t_{\rm{hidden}}$ at certain CSM density. Above this maximum CSM density, the photons in the shock can escape from the shock after the shock radius exceeds the photosphere radius in the SN ejecta, which will be observed as a normal Type IIn SN. This gives us the maximum value for $t_{\rm{hidden}}$ as $t_{\rm{hidden}} = 37.7$ days in the case of the CSM disk with $\theta_{0}=\arcsin (0.1) \sim 5.7^{\circ}$, which is achieved when $\dot{M}_{\rm{disk}}= 1.3 \times 10^{-2}$ M$_{\odot}$ yr$^{-1}$. 

In addition, the value of $t_{\rm{hidden}}$ becomes shorter if we consider the cases with larger $\theta_{0}$ or the viewing angle effects. In short, it is difficult to create such bright Type II SNe with this scenario (a normal SN plus aspherical CSM). As we have discussed above, such systems can explain rather normal Type IIP SNe with a bump in their LC.

%%% Table 1 %%%%%
\begin{table*}
\caption{Values of $t_{\rm{hidden}}$ (days) for the CSM disk with $\theta_{0}=\arcsin (0.1) \sim 5.7^{\circ}$ and various values of $\dot{M}_{\rm{disk}}$.}
  \begin{tabular}{|c||l|l|l|l|l|} \hline
    $\dot{M}_{\rm{disk}}$ & 1.0e-1.0 & 1.0e-2.0 & 1.0e-3.0 & 1.0e-4.0 \\ \hline \hline
    w/o the effects & 50.7 & 35.8 & 27.7 & 21.5 \\ \hline
    w/ the effects & 52.8 & 36.6 & 28.3 & 21.5 \\ \hline
  \end{tabular}
\end{table*}

It is important to observationally detect the system at these early (before $t_{\rm{hidden}}$) and late (after $t_{\rm{sn}}$) phases for confirming the existence of interacting SNe with a CSM disk. Especially the sudden luminosity rise with the spectral changing is an important hint for the CSM geometry. It is also important to observe the late phase of the system. This enables us to detect both of the pure $^{56}$Co decay component and the CSM interaction component. In the tail phase, the luminosity is powered by the CSM interaction and the radio active decay of $^{56}$Co. In the case with $\theta_{0}=\arcsin (0.1) \sim 5.7^{\circ}$, the contribution by the 0.032 M$_{\odot}$ of $^{56}$Co is dominant until 568.8, 375.6, 199.1 and 135.4 days after explosion for $\dot{M}_{\rm{disk}}=10^{-4}$, $10^{-3}$, $10^{-2}$ and $10^{-1}$ M$_{\odot}$ yr$^{-1}$, respectively. Here, we use the method to evaluate the luminosity from $^{56}$Co \citep[][]{Hamuy2003}. Then, the contribution from the CSM interaction becomes dominant. Therefore, we can get information on the Ni mass and the CSM interaction from observations in the early part and later part of the tail phases, respectively. As a future work, we will statistically investigate these early and late phase of Type IIn SNe using archival data in the literature.

\section{Conclusions}

We have calculated light curves for a Type IIP SN interacting with a CSM disk when viewed from the polar direction, which have been suggested as an explanation of some unusual Type II SNe, e.g., the so-called ``impossible'' SN, iPTF14hls. We have adopted the following values for the input SN: $E_{\rm{sn}} = 1 \times 10^{51}$ erg, $M_{\rm{ej}} = 10$ M$_{\odot}$, $R_{\rm{p}} = 500$ R$_{\odot}$ and $M_{\rm{Ni}}=0.032$ M$_{\odot}$. We have used the modified version of the shock interaction model of \citet{Moriya2013} embedded within the Type IIP SN model of \citet{Kasen2009}. This model has taken into account the effects of the ionization and recombination in the SN ejecta so that we could investigate the radiative processes in the systems described, e.g., in \citet{Smith2015}. It should be noted that our calculations cannot predict quantitative numbers because we have not considered the viewing angle effects and the geometrical effects of the CSM disk and because we also adopt several simplifications in the model. For example, we have assumed that we do not see narrow lines in the spectrum until the interaction exceeds the photosphere and that the SN ejecta is homologously expanding, which slightly change our derived values. For quantitative discussions, we need to conduct more detailed calculations than our calculations, i.e., multi-dimensional radiation hydrodynamics simulations, taking into account the ionization and recombination of hydrogen.

We have demonstrated that they show three phases with different photometric and spectroscopic properties, following the change of the main energy source.
(Phase~1): For first few tens days, the main energy source is the thermal energy of the SN ejecta deposited by the core-bounced shock and the synthesized radioactive elements. The energy created by the CSM interaction is injected deep into the SN ejecta, most of which is lost by the expansion cooling within the SN ejecta. In this term, the system shows similar photospheric and spectroscopic properties with those of normal Type II SNe with P-Cygni lines. 
(Phase~2): Once the CSM interaction overtakes the photosphere in the SN ejecta, the interacting region gets exposed to an observer. Then, the radiation from the system becomes a mixture of the radiation from the SN ejecta and the CSM interaction, showing high luminosity and some interaction features in the spectrum such as narrow lines. 
(Phase~3): Finally, the system becomes, $\sim 100$ days after explosion, powered by the CSM interaction and the radioactive decay because of the exhaustion of the initial thermal energy in the SN ejecta.

From the calculations, we have concluded that a bright Type II SN heated by a hidden CSM interaction is difficult to be realized. The SN ejecta can hide the interaction for 37.7 days at maximum in our simple calculations for the systems with the CSM disk with $\theta_{0}=\arcsin (0.1) \sim 5.7^{\circ}$, which highly depends on the values of $\theta_{0}$. It is also noted that this value can be changed if we take into account the viewing angle effects and the geometrical effects of the CSM disk or calculate it without using assumptions/simplifications we used in our light curve model.

We have also examined the effects of the disk parameters on their light curves, using the values of $\dot{M}_{\rm{disk}}=10^{-4}$ to $10^{-1}$ M$_{\odot}$ yr$^{-1}$ and $\theta_{0} =5$ to 45$^{\circ}$. We found that the luminosity in Phase~1 is sensitive to the opening angle of the CSM disk for a certain total mass of the CSM, while the luminosity in Phase~2 is roughly determined by the total mass of the CSM rather than the opening angle. Thus, the luminosity ratio between in the Phase~1 and in Phase~2 has information on the opening angle of the CSM disk. From this result, we have proposed the importance of early phase observations of interacting SNe with photometry and spectroscopy for investigating their CSM geometries. Here, we have just demonstrated the importance of early phase observations of the interacting SNe for investigating their CSM geometries with qualitative discussions using simplified models.

\section*{Acknowledgements}
We thank the anonymous referee for the comments, which improved this paper. We also would like to thank A. Suzuki and J. P. Anderson for useful discussions. T.N. is supported by Japan Society for the Promotion of Science (JSPS) Overseas Research Fellowship. This work has been supported by JSPS through KAKENHI grant 17H02864, 18H04585, 18H05223, 20H00174, and 20H04737 (K.M.) and 19J14158 (R.O.).

Data availability: The data underlying this article are available in the article.

%%%%%%%%%%%%%%%%%%%%%%%%%%%%%%%%%%%%%%%%%%%%%%%%%%



%\bibliographystyle{mnras}

\begin{thebibliography}{99}
\bibitem[\protect\citeauthoryear{Anderson}{2019}]{Anderson2019} Anderson J.~P., 2019, A\&A, 628, A7
\bibitem[\protect\citeauthoryear{Andrews, et al.}{2017}]{Andrews2017} Andrews J.~E., Smith N., McCully C., Fox O.~D., Valenti S., Howell D.~A., 2017, MNRAS, 471, 4047
\bibitem[\protect\citeauthoryear{Andrews \& Smith}{2018}]{Andrews2018} Andrews J.~E., Smith N., 2018, MNRAS, 477, 74
\bibitem[\protect\citeauthoryear{Andrews, et al.}{2019}]{Andrews2019} Andrews J.~E., et al., 2019, ApJ, 885, 43
\bibitem[\protect\citeauthoryear{Arcavi, et al.}{2017}]{Arcavi2017} Arcavi I., et al., 2017, Natur, 551, 210
\bibitem[\protect\citeauthoryear{Arnett \& Meakin}{2011}]{Arnett2011} Arnett W.~D., Meakin C., 2011, ApJ, 741, 33
\bibitem[\protect\citeauthoryear{Boian \& Groh}{2020}]{Boian2020} Boian I., Groh J.~H., 2020, arXiv, arXiv:2001.07651
\bibitem[\protect\citeauthoryear{Chevalier}{2012}]{Chevalier2012} Chevalier R.~A., 2012, ApJL, 752, L2
\bibitem[\protect\citeauthoryear{Elias-Rosa, et al.}{2016}]{Elias-Rosa2016} Elias-Rosa N., et al., 2016, MNRAS, 463, 3894
\bibitem[\protect\citeauthoryear{Filippenko}{1997}]{Filippenko1997} Filippenko A.~V., 1997, ARA\&A, 35, 309
\bibitem[\protect\citeauthoryear{F{\"o}rster, et al.}{2018}]{Forster2018} F{\"o}rster F., et al., 2018, NatAs, 2, 808
\bibitem[\protect\citeauthoryear{Fransson, et al.}{2002}]{Fransson2002} Fransson C., et al., 2002, ApJ, 572, 350
\bibitem[\protect\citeauthoryear{Fransson, et al.}{2014}]{Fransson2014} Fransson C., et al., 2014, ApJ, 797, 118
\bibitem[\protect\citeauthoryear{Fraser, et al.}{2013}]{Fraser2013} Fraser M., et al., 2013, ApJL, 779, L8
\bibitem[\protect\citeauthoryear{Fuller}{2017}]{Fuller2017} Fuller J., 2017, MNRAS, 470, 1642
\bibitem[\protect\citeauthoryear{Hamuy}{2003}]{Hamuy2003} Hamuy M., 2003, ApJ, 582, 905
\bibitem[\protect\citeauthoryear{Hoffman, et al.}{2008}]{Hoffman2008} Hoffman J.~L., Leonard D.~C., Chornock R., Filippenko A.~V., Barth A.~J., Matheson T., 2008, ApJ, 688, 1186
\bibitem[\protect\citeauthoryear{Humphreys \& Davidson}{1994}]{Humphreys1994} Humphreys R.~M., Davidson K., 1994, PASP, 106, 1025
\bibitem[\protect\citeauthoryear{Jerkstrand, Maeda, \& Kawabata}{2020}]{Jerkstrand2020} Jerkstrand A., Maeda K., Kawabata K.~S., 2020, Sci, 367, 415
\bibitem[\protect\citeauthoryear{Kasen \& Woosley}{2009}]{Kasen2009} Kasen D., Woosley S.~E., 2009, ApJ, 703, 2205
\bibitem[\protect\citeauthoryear{Katsuda, et al.}{2016}]{Katsuda2016} Katsuda S., et al., 2016, ApJ, 832, 194
\bibitem[\protect\citeauthoryear{Khazov, et al.}{2016}]{Khazov2016} Khazov D., et al., 2016, ApJ, 818, 3
\bibitem[\protect\citeauthoryear{Kiewe, et al.}{2012}]{Kiewe2012} Kiewe M., et al., 2012, ApJ, 744, 10
\bibitem[\protect\citeauthoryear{Kilpatrick, et al.}{2018}]{Kilpatrick2018} Kilpatrick C.~D., et al., 2018, MNRAS, 473, 4805
\bibitem[\protect\citeauthoryear{Kurf{\"u}rst \& Krti{\v{c}}ka}{2019}]{Kurfurst2019} Kurf{\"u}rst P., Krti{\v{c}}ka J., 2019, A\&A, 625, A24
\bibitem[\protect\citeauthoryear{Langer, Garc{\'\i}a-Segura \& Mac Low}{1999}]{Langer1999} Langer N., Garc{\'\i}a-Segura G., Mac Low M.-M., 1999, ApJL, 520, L49
\bibitem[\protect\citeauthoryear{Leonard, et al.}{2000}]{Leonard2000} Leonard D.~C., Filippenko A.~V., Barth A.~J., Matheson T., 2000, ApJ, 536, 239
\bibitem[\protect\citeauthoryear{Levesque, et al.}{2014}]{Levesque2014} Levesque E.~M., Stringfellow G.~S., Ginsburg A.~G., Bally J., Keeney B.~A., 2014, AJ, 147, 23
\bibitem[\protect\citeauthoryear{Matzner \& McKee}{1999}]{Matzner1999} Matzner C.~D., McKee C.~F., 1999, ApJ, 510, 379
\bibitem[\protect\citeauthoryear{Mauerhan, et al.}{2013}]{Mauerhan2013} Mauerhan J.~C., et al., 2013, MNRAS, 430, 1801
\bibitem[\protect\citeauthoryear{Mauerhan, et al.}{2014}]{Mauerhan2014} Mauerhan J., et al., 2014, MNRAS, 442, 1166
\bibitem[\protect\citeauthoryear{McDowell, Duffell \& Kasen}{2018}]{McDowell2018} McDowell A.~T., Duffell P.~C., Kasen D., 2018, ApJ, 856, 29
\bibitem[\protect\citeauthoryear{Moriya, et al.}{2013}]{Moriya2013} Moriya T.~J., Maeda K., Taddia F., Sollerman J., Blinnikov S.~I., Sorokina E.~I., 2013, MNRAS, 435, 1520
\bibitem[\protect\citeauthoryear{Moriya, et al.}{2018}]{Moriya2018} Moriya T.~J., F{\"o}rster F., Yoon S.-C., Gr{\"a}fener G., Blinnikov S.~I., 2018, MNRAS, 476, 2840
%\bibitem[\protect\citeauthoryear{Nakaoka, et al.}{2018}]{Nakaoka2018} Nakaoka T., et al., 2018, ApJ, 859, 78
\bibitem[\protect\citeauthoryear{Ofek, et al.}{2014}]{Ofek2014} Ofek E.~O., et al., 2014, ApJ, 789, 104
\bibitem[\protect\citeauthoryear{Pastorello, et al.}{2013}]{Pastorello2013} Pastorello A., et al., 2013, ApJ, 767, 1
\bibitem[\protect\citeauthoryear{Pastorello, et al.}{2018}]{Pastorello2018} Pastorello A., et al., 2018, MNRAS, 474, 197
\bibitem[\protect\citeauthoryear{Pastorello, et al.}{2019}]{Pastorello2019} Pastorello A., et al., 2019, A\&A, 628, A93
\bibitem[\protect\citeauthoryear{Patat, et al.}{2011}]{Patat2011} Patat F., Taubenberger S., Benetti S., Pastorello A., Harutyunyan A., 2011, A\&A, 527, L6
\bibitem[\protect\citeauthoryear{Quataert \& Shiode}{2012}]{Quataert2012} Quataert E., Shiode J., 2012, MNRAS, 423, L92
\bibitem[\protect\citeauthoryear{Quataert, et al.}{2016}]{Quataert2016} Quataert E., Fern{\'a}ndez R., Kasen D., Klion H., Paxton B., 2016, MNRAS, 458, 1214
\bibitem[\protect\citeauthoryear{Reilly, et al.}{2017}]{Reilly2017} Reilly E., et al., 2017, MNRAS, 470, 1491
\bibitem[\protect\citeauthoryear{Shiode \& Quataert}{2014}]{Shiode2014} Shiode J.~H., Quataert E., 2014, ApJ, 780, 96
\bibitem[\protect\citeauthoryear{Smartt, et al.}{2009}]{Smartt2009} Smartt S.~J., Eldridge J.~J., Crockett R.~M., Maund J.~R., 2009, MNRAS, 395, 1409
\bibitem[\protect\citeauthoryear{Smith, et al.}{2010}]{Smith2010} Smith N., Chornock R., Silverman J.~M., Filippenko A.~V., Foley R.~J., 2010, ApJ, 709, 856
\bibitem[\protect\citeauthoryear{Smith \& Arnett}{2014}]{Smith2014} Smith N., Arnett W.~D., 2014, ApJ, 785, 82
\bibitem[\protect\citeauthoryear{Smith, et al.}{2015}]{Smith2015} Smith N., et al., 2015, MNRAS, 449, 1876
\bibitem[\protect\citeauthoryear{Smith}{2017}]{Smith2017} Smith N., 2017, hsn..book, 403, hsn..book
\bibitem[\protect\citeauthoryear{Soker \& Kashi}{2013}]{Soker2013} Soker N., Kashi A., 2013, ApJL, 764, L6
\bibitem[\protect\citeauthoryear{Soumagnac, et al.}{2019}]{Soumagnac2019} Soumagnac M.~T., et al., 2019, ApJ, 872, 141
\bibitem[\protect\citeauthoryear{Soumagnac, et al.}{2020}]{Soumagnac2020} Soumagnac M.~T., et al., 2020, arXiv, arXiv:2001.05518
\bibitem[\protect\citeauthoryear{Suzuki, Moriya \& Takiwaki}{2019}]{Suzuki2019} Suzuki A., Moriya T.~J., Takiwaki T., 2019, ApJ, 887, 249
\bibitem[\protect\citeauthoryear{Taddia, et al.}{2013}]{Taddia2013} Taddia F., et al., 2013, A\&A, 555, A10
\bibitem[\protect\citeauthoryear{Th{\"o}ne, et al.}{2017}]{Thone2017} Th{\"o}ne C.~C., et al., 2017, A\&A, 599, A129
\bibitem[\protect\citeauthoryear{Wang, et al.}{2001}]{Wang2001} Wang L., Howell D.~A., H{\"o}flich P., Wheeler J.~C., 2001, ApJ, 550, 1030
\bibitem[\protect\citeauthoryear{Woosley \& Heger}{2015}]{Woosley2015} Woosley S.~E., Heger A., 2015, ApJ, 810, 34
\bibitem[\protect\citeauthoryear{Yaron, et al.}{2017}]{Yaron2017} Yaron O., et al., 2017, NatPh, 13, 510
\bibitem[\protect\citeauthoryear{Yoon \& Cantiello}{2010}]{Yoon2010} Yoon S.-C., Cantiello M., 2010, ApJL, 717, L62
\end{thebibliography}


%%%%%%%%%%%%%%%%%%%%%%%%%%%%%%%%%%%%%%%%%%%%%%%%%%

%%%%%%%%%%%%%%%%% APPENDICES %%%%%%%%%%%%%%%%%%%%%

\appendix
\section{Radiation from a expanding gas shell}

\begin{figure}
	\includegraphics[width=\columnwidth]{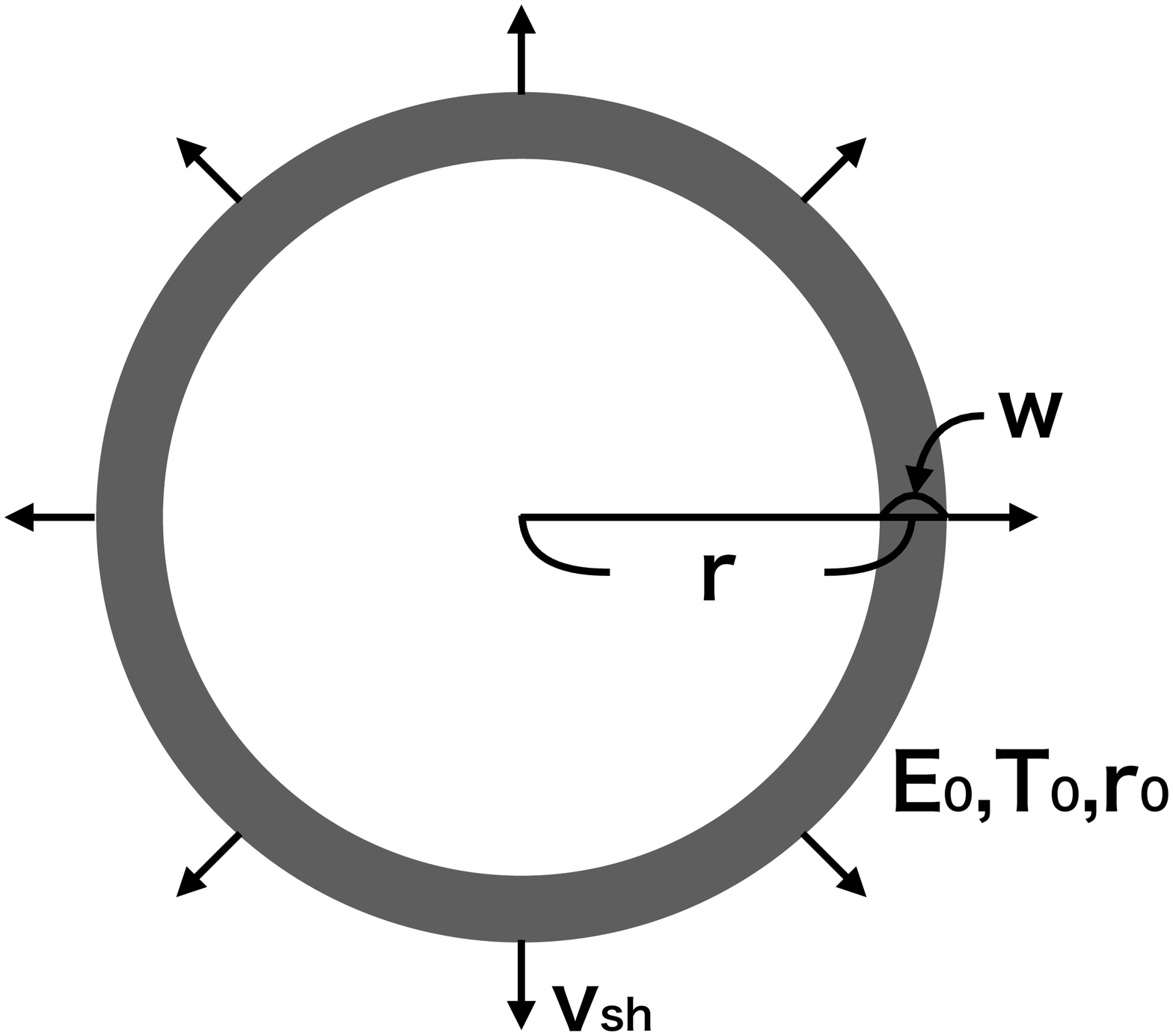}
    \caption{Schematic picture of the expanding shell of fully ionized gas}
\end{figure}

We consider radiation from a fully ionized gas shell that is expanding with a constant velocity, $v_{\rm{shell}}$, while whose internal energy leaking out through radiative diffusion processes (Fig.~A1). The initial energy, temperature and radius of the shell are denoted as $E_{0}$, $T_{0}$ and $r_{0}$, respectively. The width of the shell ($w$) is assumed to be proportional to the radius of the shell as $w = \alpha \cdot r$. The radius of the shell at time $t$ can be expresses as follows:
\begin{equation}
    r = r_{0} + v_{\rm{shell}}t = r_{0} \left( 1 + \frac{v_{\rm{shell}}}{r_{0}}t \right) = r_{0} \left( 1 + \frac{t}{t_{\rm{dyn}}(0)} \right).
\end{equation}
Here, we define the dynamical time ($t_{\rm{dyn}}$) and the diffusion time ($t_{\rm{diff}}$) of the shell as follows:
\begin{eqnarray}
t_{\rm{dyn}}(t) &=& \frac{r}{v_{\rm{shell}}},\\
t_{\rm{diff}} (t) &=& \frac{\tau w}{c}=\frac{\kappa_{\rm{es}} \cdot \frac{M}{V} \cdot w \cdot w}{c} = \frac{\kappa_{\rm{es}} \alpha M}{4 \pi c r_{0}} \frac{1}{\left( 1+\frac{t}{t_{\rm{dyn}}(0)} \right)} \nonumber\\
&=& t_{\rm{diff}} (0)\frac{1}{\left( 1+\frac{t}{t_{\rm{dyn}}(0)} \right)},
\end{eqnarray}
where $\tau$, $M$ and $V=4 \pi r^{2} w = 4 \pi \alpha r^{3}$ are the optical depth, mass and volume of the shell. Here, we consider the situation that the energy and pressure of radiation are dominant compared with those of ions and electrons. Thus, the energy ($E$) and pressure ($P$) are expressed as follows:
\begin{eqnarray}
E &=& a T^{4} V = 4 \pi \alpha a T^{4} r^{3},\\
P &=& \frac{1}{3} a T^{4},
\end{eqnarray}
where $a$ is the radiation constant as $a=4\sigma_{\rm{SB}} /c$, and $\sigma_{\rm{SB}}$ is the Stefan-Boltzmann constant.

From the first law of thermodynamics with an assumption that the thermal energy of the shell is leaking through radiative diffusion processes as $E/t_{\rm{diff}}(t)$, we get the temperature evolution as follows:
\begin{eqnarray}
&&\frac{dE}{dt} + P \frac{dV}{dt} = - \frac{E}{t_{\rm{diff}}(t)},\\
&&\therefore \frac{1}{T} \frac{dT}{dt} + \frac{1}{r} \frac{dr}{dt} = - \frac{1}{4 t_{\rm{diff}}(t)},\\
&&\therefore \frac{d}{dt} \left( \ln \left(Tr \right) \right) = - \frac{1}{4 t_{\rm{diff}}(0)} \left( 1 + \frac{t}{t_{\rm{dyn}}(0)} \right),\\
&& \therefore T = \frac{T_{0} r_{0}}{r} \exp \left[ - \frac{t}{4t_{\rm{diff}}(0)} \left( 1+ \frac{t}{2 t_{\rm{dyn}}(0)} \right) \right].
\end{eqnarray}

Therefore, we derive the luminosity of the shell as radiative diffusion loss:
\begin{eqnarray}
L &=& \frac{E}{t_{\rm{diff}}(t)} = \frac{aT^{4}V}{t_{\rm{diff}}(t)} = \frac{16 \pi^{2} a c}{\kappa_{\rm{es}}M} T^{4}r^{4},\\
&=& \frac{16 \pi^{2}ac T_{0}^{4} r_{0}^{4}}{\kappa_{\rm{es}}M} \exp \left[  -\frac{t}{t_{\rm{diff}}(0)} \left( 1 + \frac{t}{2 t_{\rm{dyn}}(0)} \right) \right],\\
&=& \frac{E_{0}}{t_{\rm{diff}}(0)} \exp \left[  -\frac{t}{t_{\rm{diff}}(0)} \left( 1 + \frac{t}{2 t_{\rm{dyn}}(0)} \right) \right].
\end{eqnarray}

%%%%%%%%%%%%%%%%%%%%%%%%%%%%%%%%%%%%%%%%%%%%%%%%%%

%\begin{figure}
	%\includegraphics[width=\columnwidth]{example}
    %\caption{This is an example figure. Captions appear below each figure.
	%Give enough detail for the reader to understand what they're looking at,
	%but leave detailed discussion to the main body of the text.}
    %\label{fig:example_figure}
%\end{figure}

% Don't change these lines
\bsp	% typesetting comment
\label{lastpage}
\end{document}